\documentclass[iop,apj]{emulateapj}

\usepackage{amsmath,amssymb,amstext}

\usepackage[breaklinks,colorlinks,citecolor=blue,linkcolor=magenta]{hyperref}

\usepackage[all]{hypcap} 

\usepackage{aas_macros}
\usepackage{natbib}
\usepackage[usenames,dvipsnames]{color}

\newcommand{\tff}{t_{\mbox{\scriptsize{\em ff}}}}

\shorttitle{Velocity Dispersion-Size Relation in Collapsing Clouds.}
\shortauthors{Ib\'a\~nez-Mej\'{\i}a et. al.}

\begin{document}

\title{Gravitational contraction versus supernova driving and the origin of the velocity dispersion-size relation in molecular clouds}

\author{Juan C. Ib\'a\~nez-Mej\'{\i}a\altaffilmark{1,2}, Mordecai-Mark Mac Low\altaffilmark{2,1}, Ralf S. Klessen\altaffilmark{1} and Christian Baczynski\altaffilmark{1, 3}}

\altaffiltext{1}{Institut f\"ur Theoretische Astrophysik, Zentrum f\"ur Astronomie der Universit\"at Heidelberg, Albert-Ueberle-Str. 2, 69120 Heidelberg, Germany, \email{jibanez@zah.uni-heidelberg.de}}

\altaffiltext{2}{American Museum of Natural History, 79th St. at Central Park West, New York, NY 10024, USA}

\altaffiltext{3}{School of Physics and Astronomy, University of St.\ Andrews, North Haugh, St Andrews KY16 9SS, UK}

\begin{abstract}

Molecular cloud observations show that clouds have non-thermal velocity dispersions that scale with the cloud size as $\sigma\propto R^{1/2}$ at constant surface density, and for varying surface density scale with both the cloud's size and surface density, $\sigma^2 \propto R \Sigma $.
The energy source driving these chaotic motions remains poorly understood.
We describe the velocity dispersions observed in a cloud population formed in a numerical simulation of a magnetized, stratified, supernova-driven, interstellar medium, including diffuse heating and radiative cooling, before and after we include the effects of the self-gravity of the gas.
We compare the relationships between velocity dispersion, size, and surface density measured in the simulated cloud population to those found in observations of Galactic molecular clouds.
Our simulations prior to the onset of self-gravity suggest that external supernova explosions alone do not drive turbulent motions of the observed magnitudes within dense clouds. On the other hand, self-gravity induces non-thermal motions as gravitationally bound clouds begin to collapse in our model, approaching the observed relations between velocity dispersion, size, and surface density.
Energy conservation suggests that the observed behavior is consistent with the kinetic energy being proportional to the gravitational energy. 
However, the clouds in our model show no sign of reaching a stable equilibrium state at any time, even for strongly magnetized clouds.
We conclude that gravitationally bound molecular clouds are always in a state of gravitational contraction and their properties are a natural result of this chaotic collapse.  In order to agree with observed star formation efficiencies, this process must be terminated by the early destruction of the clouds, presumably from internal stellar feedback.

\end{abstract}

\maketitle

\section{Introduction}
\label{sec:Introduction}

Understanding what regulates molecular cloud (MC) properties is key to understanding their evolution and role in the star formation process.
Four decades ago, molecular line observations of dense interstellar clouds revealed that clouds have internal velocity gradients far larger than expected from thermal velocities \citep{Zuckerman1974RadioMolecules}.
These fast turbulent motions were first interpreted as signatures of gravitational collapse \citep{Goldreich1974MolecularClouds}.
However if the observed MCs were collapsing in a free-fall time, the expected star formation rate would be an order of magnitude larger than the observed rate \citep{Zuckerman1974RadioMolecules}.
In reality the star formation process is controlled by a non-linear combination of self-gravity, turbulence, magnetic fields, radiation, and gas heating and cooling \citep[e.g.][]{MacLow2004ControlTurbulence, Klessen2014PhysicalMedium}. 
How these processes come together to regulate the formation, evolution, and collapse of MCs remains a subject of active research \citep[][and references therein]{Dobbs2014PPVI}.
Idealized simulations of artificially driven turbulence in isolated MCs has provided the foundations for present analytical star formation models \citep{Krumholz2005AGalaxies, Padoan2012AFormation,Federrath2012THEOBSERVATIONS}.
However it remains unknown if these simulations accurately capture the processes dominating real MC properties, and therefore represent real star formation relations.
In this paper we explore the interaction between self-gravity and turbulence for a simulated cloud population formed in a kiloparsec-scale, magnetized, supernova (SN) driven, turbulent, interstellar medium (ISM), and compare the properties of the simulated clouds with the properties of observed MCs in the Galaxy.

In his seminal paper, \citet[][hereafter L81]{Larson1981} proposed two scaling relations, now known as ``Larson's relations,'' that related the volume density, size, and velocity dispersion of MCs.
This work provided a fundamental insight into cloud dynamics.
These relations have been extensively re-examined and are now believed to have the form \citep{Solomon1987MassClouds, Heyer2009RE-EXAMININGCLOUDS, Falgarone2009IntermittencyScale}:
\begin{eqnarray}
\sigma & \propto R^{0.5}  \\
\rho  & \propto R^{-1.1}  
\end{eqnarray}
The first of these relations tells us that clouds are turbulent structures, and is often interpreted as occurring due to a Kolmogorov-like cascade for supersonic turbulent motions \citep{Larson1981, Kritsuk2013ALaws, Klessen2014PhysicalMedium, Padoan2015OriginTurbulence}.
The second relation implies a constant column density for all clouds.

\citet{Heyer2009RE-EXAMININGCLOUDS} re-examined Larson's relations using observations with $^{13}$CO, a lower opacity tracer than the original $^{12}$CO, that reduces velocity crowding and allows for a more direct measurement of molecular column density.
He found that, contrary to Larson's constant surface density result, clouds exhibit a large dynamical range in surface densities.
He was able to extend Larson's velocity dispersion-size relation to include the variation in the cloud's surface density,
\begin{equation}
	\sigma = K \Sigma^{1/2} R^{1/2}.
\end{equation}
In his study, \citet{Heyer2009RE-EXAMININGCLOUDS} concluded that this dependence reflects clouds in a state of virial equilibrium, so that the constant of proportionality is 
\begin{equation} K = (\pi G / 5)^{1/2} \label{eq:virial_eq}. \end{equation} 
However \citet{Ballesteros-Paredes2011GravityRelation} noted that the velocity dispersion in clouds collapsing at the free-fall velocity differs from that of clouds in virial equilibrium by a factor of only $\sqrt{2}$, making it difficult to differentiate between these two states on the basis of velocity dispersion alone. 

The energy source for MC turbulence remains controversial, although many candidates have been proposed. In particular the question remains unanswered of whether turbulence is driven from the inside, by protostellar outflows \citep{Li2006ClusterTurbulence, Banerjee2006OutflowsCores, Banerjee2007ProtostellarClouds, Nakamura2014CONFRONTINGOBSERVATIONS, Federrath2014MODELINGFORMATION, Offner2015IMPACTTURBULENCE}, expanding HII regions \citep{Dale2012IonizingClusters, Walch2012DispersalRadiation, Dale2014}, or stellar winds and internal SNe \citep{Iffrig2015MutualClouds}, or driven from the outside, by external SN explosions \citep{MacLow2004ControlTurbulence, Walch2015TheClouds}, colliding flows or tidal forces \citep{VazquezSemadeni2006MolecularFormation, Ballesteros-Paredes2009TidalTaurus}, or accretion and collapse \citep{Klessen2010Accretion-drivenDisks, Ballesteros-Paredes2011GravityRelation, Goldbaum2011THEACCRETION, Heitsch2013GRAVITATIONALFILAMENTS, Traficante2015}.

Observations tell us that the turbulent energy is mostly contained at the largest scales in MCs \citep{MacLow2000CharacterizingTurbulence,Brunt2003LargeScaleClouds, Brunt2009TurbulentClouds} which makes it difficult for internal sources to drive the turbulence.
Combined SN explosions in the field have been shown to drive turbulence at scales of 100--200~pc \citep{Joung2006TurbulentMedium,Avillez2007TheSimulations} and have been suggested to be regular and energetic enough to maintain turbulence at all scales \citep{MacLow2004ControlTurbulence}.
Self-gravity has also been argued to be one of the main drivers of turbulence \citep{Elmegreen1993,VazquezSemadeni2006MolecularFormation,VazquezSemadeni2007MolecularConditions, Elmegreen2007}. 
For example gravitational collapse of a hierarchically structured cloud could drive seemingly random velocities in agreement with observations.
Self-gravity can also result in accretion driven turbulence as material falls onto a MC \citep{Vazquez-Semadeni2008TheCase, Heitsch2008RapidMovies, Klessen2010Accretion-drivenDisks, Ballesteros-Paredes2011GravityRelation, Goldbaum2011THEACCRETION, Heitsch2013GRAVITATIONALFILAMENTS,Traficante2015}. 

This paper is organized as follows:
in Section 2 we describe the simulations, our cloud identification algorithm, and the method of analysis.
In Section 3 we present the properties derived for the simulated clouds and how they compare with observations. 
We discuss the implications of our results in Section 4, and summarize our results in Section 5.

\section{Methods and Simulations}
\label{sec:methods}

\subsection{Stratified Box Simulation}
\label{sec:StratBox}

We present and analyze results from three-dimensional numerical simulations of self-gravitating, magnetized, SN-driven turbulence in the ISM.
These simulations correspond to a direct extension of the stratified box models by \citet{Joung2006TurbulentMedium, Joung2009}, and \citet{Hill2012VerticalMedium}, now including gas self-gravity and higher resolution in dense regions. 

The simulation uses a grid in the shape of an elongated box of size $1 \times 1 \times 40 \, \rm{kpc}^{3}$, centered on  
the galactic midplane. We use periodic boundary conditions in the horizontal directions, and outflow boundary conditions at the top and bottom of the box.

A static disk gravitational potential represents the gravitational influence of dark matter and already existing stars in and above the disk.  Near the disk, the potential follows a modified version of the solar neighborhood potential derived by \citet{Kuijken1989TheSun}, transitioning to the inner halo potential of \citet{Dehnen1998MassWay} at $|z| \geq 4 \rm{\, kpc}$. At heights above $|z| \geq 7.5 \rm{\, kpc}$, there is a smooth transition to the outer halo potential of \citet[hereafter NFW]{Navarro1996TheHalos}. The gravitational acceleration resulting is
\begin{eqnarray}
\begin{split}
	g(z)  & = - \frac{ a_{1} z } {\sqrt{ z^{2} + z_{0}^2}} - a_{2}z + a_{3} z |z|,   & |z| \leq \mbox{ 7.5 kpc} \\
              & = - \frac{4}{3} G \pi \rho_h  z,   & |z| > \mbox{ 7.5 kpc}
\end{split}
\end{eqnarray}
where $a_{1} = 1.42 \times 10^{-3} \rm{\, kpc \, Myr^{-2}}$, $a_{2} = 5.49 \times 10^{-4} \rm{\, Myr^{-2}}$, $a_{3} = 5 \times 10^{-5} \rm{ \, kpc^{-1} \, Myr^{-2}}$ and $z_{0} = 0.18 \rm{\, kpc}$.
For the NFW potential, $\rho_h$ is given by 
\begin{eqnarray}
	\rho_h = \rho_{s} \frac{r_{s}}{|z|}\left( 1 + \frac{|z|}{r_{s}} \right)^{-2},
\end{eqnarray}
where $r_{s} = 20 \, \rm{kpc}$ and $\rho_{s} = 9.2053 \times 10^{-25} \rm{\, g \, cm^{-3}}$.
%
The initial density distribution corresponds to a quasi-hydrostatic equilibrium between the pull of the static galactic gravitational potential and the stratification of an isothermal gas given by
%
\begin{align}
	\rho_i(z) =  \rho_i(0) \textrm{exp}  \left[  \right( &-a_{1} \sqrt{z^{2} + a_{3}^2} - \frac{1}{2} a_{2}  z^{2}  \\  
   &+ \frac{1}{3} a_{4} z^3 + a_{1} a_{3} \left) \frac{\rho_i(0)}{p_i(0)} \right], \nonumber  
\end{align}
where the density, temperature and pressure of the ISM at the midplane are $\rho_i(0) =  3.41 \times 10^{-24} \rm{ \, g \, cm^{-3}} $, $T_i = 1.15 \times 10^4  \text{ K}$  and $p_i(0) = 2.48 \times 10^{-12} \rm{\, g \, cm^{-1} \, s^{-2}}$.    
We use a mean mass per particle of $\mu = 1.3017 m_{\rm{H}}$ throughout the paper, assuming neutral, atomic gas with a helium fraction of 0.097 and the remaining 0.3\% in metals.

A uniform intergalactic medium (IGM) with density  $\rho_g = 1.72 \times 10^{-31} \text{ g cm}^{-3}$, temperature representative of a hot outer halo $T_g = 1.15\times 10^{6} \, \rm{K}$ and pressure $p_g = 1.28 \times 10^{-17}  \rm{\, g \,cm^{-1} \, s^{-2}}$ is included once the ISM density from hydrostatic equilibrium drops below the IGM density,  $\rho_i(z) < \rho_g$.
The total amount of gas in the simulation is scaled such that the projected surface density along the vertical direction, $\hat{z}$, is equal to the gas surface density in the solar neighborhood $\Sigma_{\odot} = 13.7$~M$_{\odot}$ \citep{vanderKruitTheGalaxies, OllingLuminousWay}.

We include a uniform magnetic field along the horizontal, $\hat{x}$, direction that decays exponentially with height, such that the initial plasma beta parameter $\beta = p 8 \pi / B^{2} = 2.5$ everywhere.
\citet{Hill2012VerticalMedium} has shown that the magnetic field naturally evolves in the simulation being advected by the fluid and getting tangled thanks to the SN turbulence. 
However, because no galactic shear is included in our simulations the large-scale dynamo necessary to maintain a strong, organized magnetic field cannot act.
Thus our simulations do underestimate the effects of organized large scale magnetic fields.
 
Discrete SN explosions drive the turbulence in the simulation.
Supernova rates are normalized to the galactic SN rate \citep{Tammann1994TheRate}:
Type Ia and core-collapse SN have rates of $6.58$ and $27.4 \, \rm{Myr^{-1} kpc^{-2}}$, respectively. 
The positions of the SN explosions are randomly located in the simulation box with a peak in the probability distribution at the midplane and an exponential decay proportional to the distance to the midplane. 
Vertical scale heights of 90~pc for core-collapse SNe and 325~pc for Type Ia SNe are assumed. 

SN explosions are treated as in \citet{Joung2006TurbulentMedium} and \citet{Hill2012VerticalMedium}: we add $10^{51} \, \rm{erg}$ of energy \citep{McKee1977ASubstrate, Ostriker1988AstrophysicalBlastwaves} to a sphere enclosing $60 \rm{\, M}_{\odot}$ centered at the SN position. 
No gas mass is added to the SN explosion. 
Clustered SNe are taken into account by assuming that three-fifths of the core-collapse SN are correlated in space and time, allowing superbubbles (SB) to form.  
In order to model the dynamics of moving OB associations, SB locations are treated as massless particles moving in a straight line with a velocity given by the bulk velocity of the gas at their birthplace with a maximum velocity of $20$~km~s$^{-1}$.
Most of the SB population moves at this maximum velocity, because there is a higher probability that particles are formed in fast moving, hot, diffuse gas due to its high volume filling fraction. 
The SN population in a SB is drawn from a random distribution $dN_{\textrm{SB}} \propto n_{*}^{-2} dn_{*}$ with lower and upper cut offs of $n_{*,min} = 7 $~SN and $n_{*,max} = 40 $~SN \citep{McKee1997TheGalaxy}. 
SB have a fixed lifetime of $t_{\rm{SB}} = 40$~Myr. 
SN explosions in a SB are injected at uniform time intervals distributed over the lifetime of the SB, $\Delta t_{\rm{SN, SB}} = t_{\rm{SB}} / dN_{\rm{SB}}$. 
The simulation is initialized with no pre-existing SB particles.
As the SB population builds up during the first 50~Myr of the simulation, the total SN rate of the simulation increases by a factor of two. 
After 50~Myr, new SBs are created at the same rate as old SBs disappear.
From this point onward, the total SN rate of the simulation remains roughly constant.

Radiative cooling is included corresponding to an optically thin plasma with Solar metallicity. 
The cooling curve is a piece-wise power law, following that of \citet{Dalgarno1972HeatingRegions}, with an electron fraction of $n_{e} / n_{\rm{H}} = 10^{-2}$ at $T \leq 2\times 10^{4} $~K, and cooling by resonance lines \citep{Sutherland1993CoolingPlasmas} for $T > 2\times 10^{4} $~K, as shown in Figure~1 of \citet{Joung2006TurbulentMedium}. 
Photoelectric heating from irradiated dust grains is the dominant heating mechanism for the cold and warm neutral medium \citep{Bakes1994TheHydrocarbons}.
The heating rate $\Gamma_{pe}$ is given by \citet{Wolfire1995TheMedium}, and is assumed to be independent of gas density.
We use a heating efficiency of $\epsilon=0.05$ and an incident interstellar far-ultraviolet radiation field, as proposed by \citet{Habing1968TheA}, with value of the normalization constant $G_{0} = 1.7$ given by \citet{Draine1978PhotoelectricGas}.
We assume the heating rate declines exponentially with height $\Gamma_{pe}(z)=\Gamma_{pe,0}e^{-z/h_{pe}}$, using a scale height of $h_{pe}=300 $~pc.

We run the stratified box simulations without self-gravity for $230 $~Myr.
During this period, SN explosions inject energy to the ISM, providing the energy to support the midplane from collapsing, and establishing the disk scale height \citep{Ostriker2010REGULATIONMODEL, Shetty2012MAXIMALLYREGULATION, Hill2012VerticalMedium, Kim2013THREE-DIMENSIONALRATES, Walch2015SILCC, Girichidis2016CR, Girichidis2016SILCC}, as well as forming dense clouds in converging flows. SN explosions are present during the entire evolution of the simulation.  Initially,
clouds form from convergent flows driven by SN shock fronts.
During the non-self-gravitating evolution of the simulation, clouds can not gravitationally collapse but are continuously shocked and pushed around by large-scale flows.
The gas naturally forms a multiphase ISM with most of the mass concentrated in the cold, dense phase while most of the volume is filled by warm and hot diffuse gas, as discussed in \citet{Hill2012VerticalMedium}.

Nested refinement regions are used to enforce high resolution in the midplane and lower resolution at high altitudes.
Resolution decreases by a factor of two at $|z| = 300$~pc, 1~kpc, 3~kpc and 10~kpc.
This refinement is static and does not react to strong shocks or gas condensations, which ensures that the bulk of the computational cost is concentrated on following the gas dynamics at the midplane.

We set the initial maximum resolution at the midplane to be 3.80~pc and run the simulation for 200 Myr, including SN feedback, static galactic gravitational potential, and magnetic fields, but no gas self-gravity. 
This establishes the vertical profile of the galactic fountain at modest computational cost.
A step by step increment of the refinement is then adopted in order to ensure a turbulent cascade has had the time to develop in our highest resolution regions.
At $200$~Myr we increase the maximum resolution to $1.90$~pc, correspondingly increasing the lower refinement levels, and run the simulation for $20$~Myr.
At $220$~Myr we include an extra resolution level using adaptive mesh refinement (AMR) spatially constrained to act in regions within z~$\leq \lvert 50 \lvert$~pc that would formally be unstable if self-gravity were included, with a maximum resolution of $0.95$~pc.
After that, at 230~Myr, gas self-gravity is turned on.
Table \ref{tab:resolution_table} shows the final state of the grid refinement at the moment we turn on self-gravity.
We do not include sink particles in these simulations but allow gas to collapse to the grid scale without any additional refinement in self-gravitating clouds.

\begin{table}
	\begin{center}
		\begin{tabular}{clc} 
			\toprule	
			resolution [pc]	& \multicolumn{1}{c}{height} & ref. type \\ 
    		\hline
			0.95 & $\qquad\qquad\;\;$ z $\leq \lvert $50$\lvert$~pc& AMR     \\
			1.90 &$\qquad\qquad\;\;$ z $\leq \lvert $300$\lvert$~pc& static  \\
			3.80 &$\lvert$300$\lvert$~pc  $<$ z $<\lvert$1$\lvert$~kpc& static  \\
			7.60 &$\;\;\lvert$1$\lvert$~kpc $<$ z $<\lvert$3$\lvert$~kpc& static \\
			15.2 &$\;\,\lvert$3$\lvert$~kpc $<$ z $<\lvert$10$\lvert$~kpc& static\\
			30.4 &$\lvert$10$\lvert$~kpc $<$ z $<\lvert$20$\lvert$~kpc& static  \\
    		\hline
		\end{tabular}
    \end{center}
   \caption{Final grid refinement at $230$~Myr of evolution, the moment at which we turn on gas self-gravity in the simulation. Nested static layers of grid refinement with an additional level of AMR ensure the bulk of the computational effort focuses on dense clouds in the midplane. \label{tab:resolution_table}}   
\end{table}

\subsection{Cloud Identification}
\label{sec:CloudCatalog}

In order to investigate the properties of individual giant MCs, we need to extract them from our simulations.
Ideally a comparison between simulations and observations would include chemistry and radiative transport in order to capture the non-equilibrium abundance of molecules and model the excitation and attenuation of molecular lines.
This is however out of the scope of this paper, so we do not include a model for chemistry, and identify clouds instead by a density threshold. 
This still allows us to investigate the dynamical properties of the clouds in our simulations.
We define our clouds as connected structures above a volume density threshold of $n_{th} = 100 \text{ cm}^{-3}$, chosen to roughly follow the region containing the observable tracer molecule CO.
In order to investigate the variation of the velocity dispersion with the size and surface density, we perform our analyses for two different density ranges within the clouds, inspired by the different density ranges traced by commonly observed molecules such as CO, CS, NH$_{3}$, N$_{2}$H$^{+}$ or HCO$^{+}$ \citep{Shirley15CriticalTracers}. 

The low density range covers number densities between $100 \, \text{ cm}^{-3} \leq n_{\rm{low}} \leq 5000 \, \text{ cm}^{-3}$.
This approximately represents the gas densities at which CO is abundant in the gas phase, and its emission is excited \citep{Draine2011Book, Klessen2014PhysicalMedium}. 
Although $^{12}$CO lines quickly saturate for typical column densities encountered in gas at number densities $\sim 200$~cm$^{-3}$, velocity gradients within the cloud reduce line overlap allowing more CO line photons to escape and be observed \citet{Shetty2011COEmission}, up to number densities of $\la 5\times 10^{3}$~cm$^{-3}$ making it a good tracer for the dynamics of molecular cloud envelopes. 
Hereafter, we refer to the structures captured by this density range as ``clouds''. 
The high density range corresponds to number densities between $5 \times 10^{3} \, \text{ cm}^{-3} \leq n_{\rm{high}} \leq 10^{5} \, \text{ cm}^{-3}$.
This density range roughly correspond to the volume densities where (1-0) transitions from CS, NH$_{3}$, N$_{2}$H$^{+}$ or HCO$^{+}$ are observed \citep{EvansII1999PhysicalFormation, Shirley15CriticalTracers}.
Hereafter, we call the structures captured by the high density tracer ``clumps''.

As an example of the need for proxies of different molecular tracers, it has been suggested that the MCs envelopes contain most of the observed CO, and that these envelopes evolve more slowly than the dense cores where stars form \citep{Elmegreen2007}.
Given that we perform our analysis on mass weighted quantities, if we included all the gas above a volume density threshold of $n_{thr} \geq 100$~cm$^{-3}$, our results would be dominated by these dense, quickly evolving cores, and thus could not be directly compared to CO observations.

We identify our clouds and clumps in three-dimensional Position-Position-Position (PPP) space rather than in the projected Position-Position-Velocity (PPV) as done in the observations.
However, previous studies of turbulent boxes show that the results for $\sigma-R$ power law relations do not vary significantly between PPP and PPV analysis \citep{Ballesteros-Paredes2002PhysicalModels,Shetty2010THERELATIONSHIPS, Beaumont2013QuantifyingSimulations}. 
A recent study by \citet{Pan2015WhatClouds} also compared the properties of GMCs in PPP and PPV space in galactic disk simulations, again concluding that both techniques seem to identify the same structures.

We compute the mass for each structure by integrating the total amount of mass within each density range, $M_{\rho} = \sum \rho_{i} \Delta x_{i}^{3} $, given the volume density  $\rho_{i}$ and the cell volume $\Delta x_{i}^{3}$ for all cells $N$ belonging to a cloud, excluding clumps within, or to a clump.
We calculate the size as the radius of a sphere equal to the volume encompassed by the lower threshold of a given density range, $\rm{R_{\rho} = (3 V_{\rho} / 4\pi  )^{1/3}}$.

In order to resolve the turbulent motions above the numerical dissipation scale, a minimum resolution of 10 cells is necessary \citep[][noting that \citet{Konstandin15HierarchicalTurbulence} already reports some numerical dissipation of turbulent modes resolved with less than 50 cells]{Kritsuk06AdaptiveTurbulence}.
We consider resolved structures those  with an effective diameter of $2 R_{\rho} = 10 \Delta x$, to ensure that their internal turbulent velocities are not significantly suppressed by numerical diffusion.
In this work, this condition corresponds to a minimum radius \footnote{We use the cloud radius in this study because most observational re-examinations of Larson's relation in the literature use this variable.} of our clouds and clumps $R_{\rho} \geq 4.8$~pc.

When self-gravity is included the relevant length scale is the Jeans length,
\begin{eqnarray}
\begin{split}
	\lambda_{j}(n, T) & = \left({\frac{15 k_{B} T}{4 \pi G \mu^{2} n}}\right)^{1/2} \\
    & = 3.31 \,\rm{pc} \left(\frac{n}{100\,\rm{cm}^{-3}}\right)^{-1/2}\left(\frac{T}{20 \, \rm{K}}\right)^{1/2},
\end{split}
\end{eqnarray}
where $k_{B}$ is the Boltzmann constant, $G$ is the gravitational constant and $\mu=1.3017 m_{H}$ is the mean mass per particle assuming neutral, atomic gas with a helium fraction of 0.097 and the remaining 0.3\% in metals. 
We resolve this length with at least 3.5 cells in the low density range gas.
This is marginally below the four cell resolution required by the \citet{Truelove1997TheHydrodynamics} criterion in order to avoid numerical fragmentation in a differentially rotating disk. Therefore the peak densities and fragmentation within our clouds and, particularly, our clumps are underestimated.
Nevertheless, we recover useful information on the velocity dispersion driven by gas self-gravity.
Detailed analysis of cloud and clump sub-structure requires higher resolution, so we defer that analysis to a future paper describing zoom-in simulations.

To obtain a velocity dispersion-size relation, we calculate the mass-weighted, one-dimensional, velocity dispersion for each density range using the three-dimensional velocity components $v_{x}, v_{y} $ and $v_{z}$, as well as the density $\rho$.  
For any observed cloud, denser gas contributes more to the observed linewidths.
The summation is done over all N zones within the desired density range to give
\begin{eqnarray}
	\sigma_{\rm{\rho},\text{1D}}^2 =\frac{1}{3} \frac{ \sum^{N}_{i} \rho_{i}   (\vec{v}_{i} - \bar{\vec{v}})^2 }{ \sum \rho_{i} },
\end{eqnarray}
where $\bar{\vec{v}}$ is the average, mass-weighted velocity summed over all zones in the cloud.
Since $\sigma_{\rho, \small {1D}}$ corresponds only to the non-thermal, turbulent velocities for a given density tracer, we compute the total velocity dispersion including the average mass-weighted sound speed, $\bar{c}_{s}$, 
\begin{eqnarray}
	\label{eq:sigma_tot}
	\sigma_{\rho, tot}^2 = \sigma_{\rho, \small{1D}}^{2} + \bar{c}_{s}^2.
\end{eqnarray}

In order to quantify the evolution of each cloud when self-gravity is included, we define the individual free-fall time for each cloud as the free-fall for the equivalent, spherically symmetric distribution of gas 
\begin{eqnarray}
\tff  = (3 \pi/32 G \bar{\rho})^{1/2}, 
\end{eqnarray}
where $\bar{\rho} $ is the average density accounting for all the mass in the cloud or clump.
Finally we compute the surface density for a given density range, as the projection of the mass on the area of a circle given by:
\begin{eqnarray}
 	\Sigma_{\rm{\rho}} = \frac{ M_{\rm{\rho}} }{  \pi R_{\rm{\rho}}^{2} },
 \end{eqnarray}
where $M_{\rho}$ is the mass within a given density range and $R_{\rho}$ is the radius computed from the volume enclosed by the lower threshold of the density range.

We are also interested in the evolution of these structures in time.
To follow this, we include tracer particles in our simulation, injecting 5 million particles around the midplane in the region $|z| \leq 50$~pc, at $t_{SG} = 0$.
We extract a cloud population at each snapshot and identify the tracer particles inside each cloud.
Finally, clouds are linked through time using the known trajectories of the tracer particles, building cloud evolutionary histories.

\section{Results}
\label{sec:Results}

\begin{figure*}
\centering 
\label{fig:StratBox_composite}
\includegraphics[width=1\textwidth]{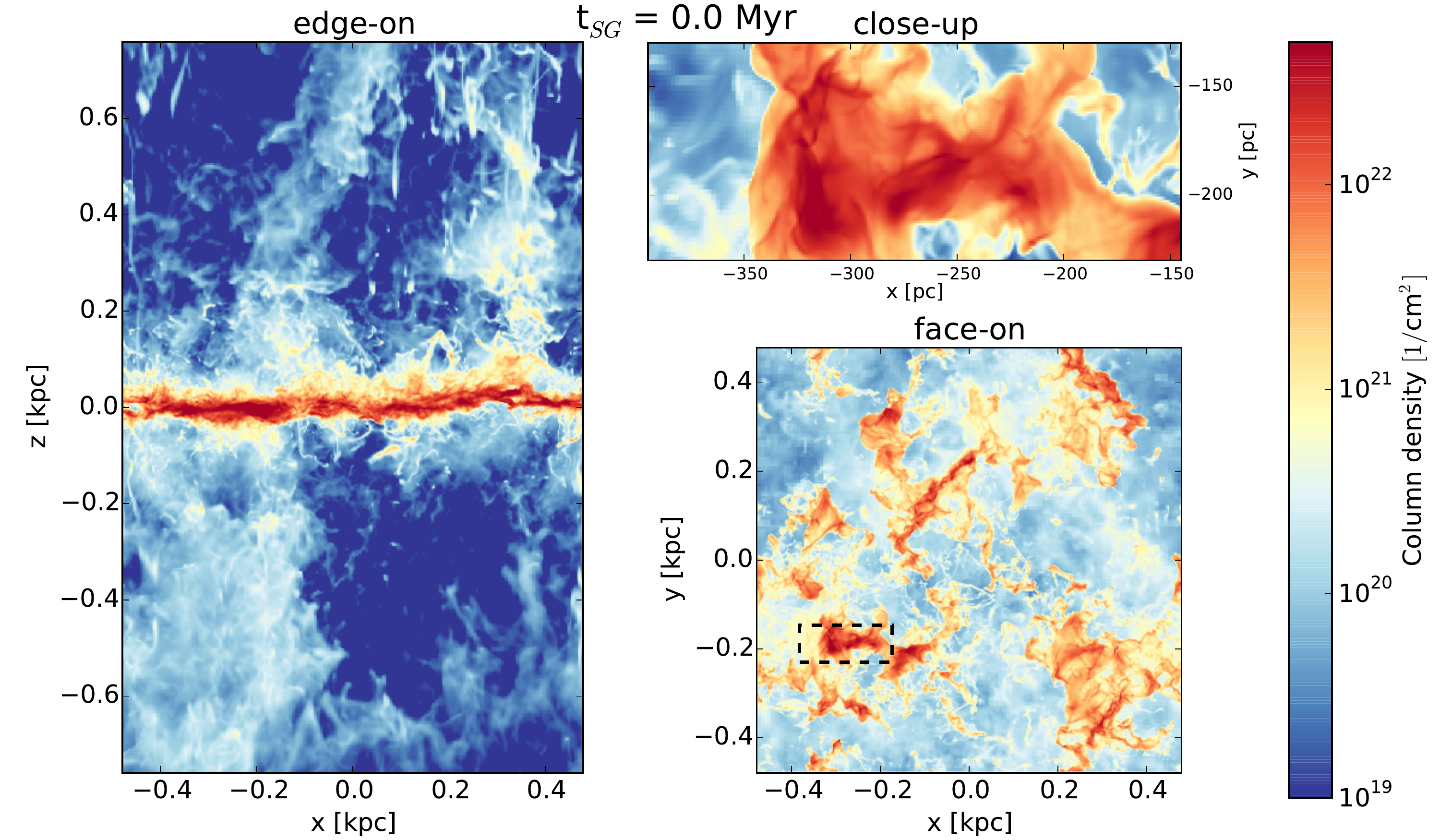} \\
\includegraphics[width=1\textwidth]{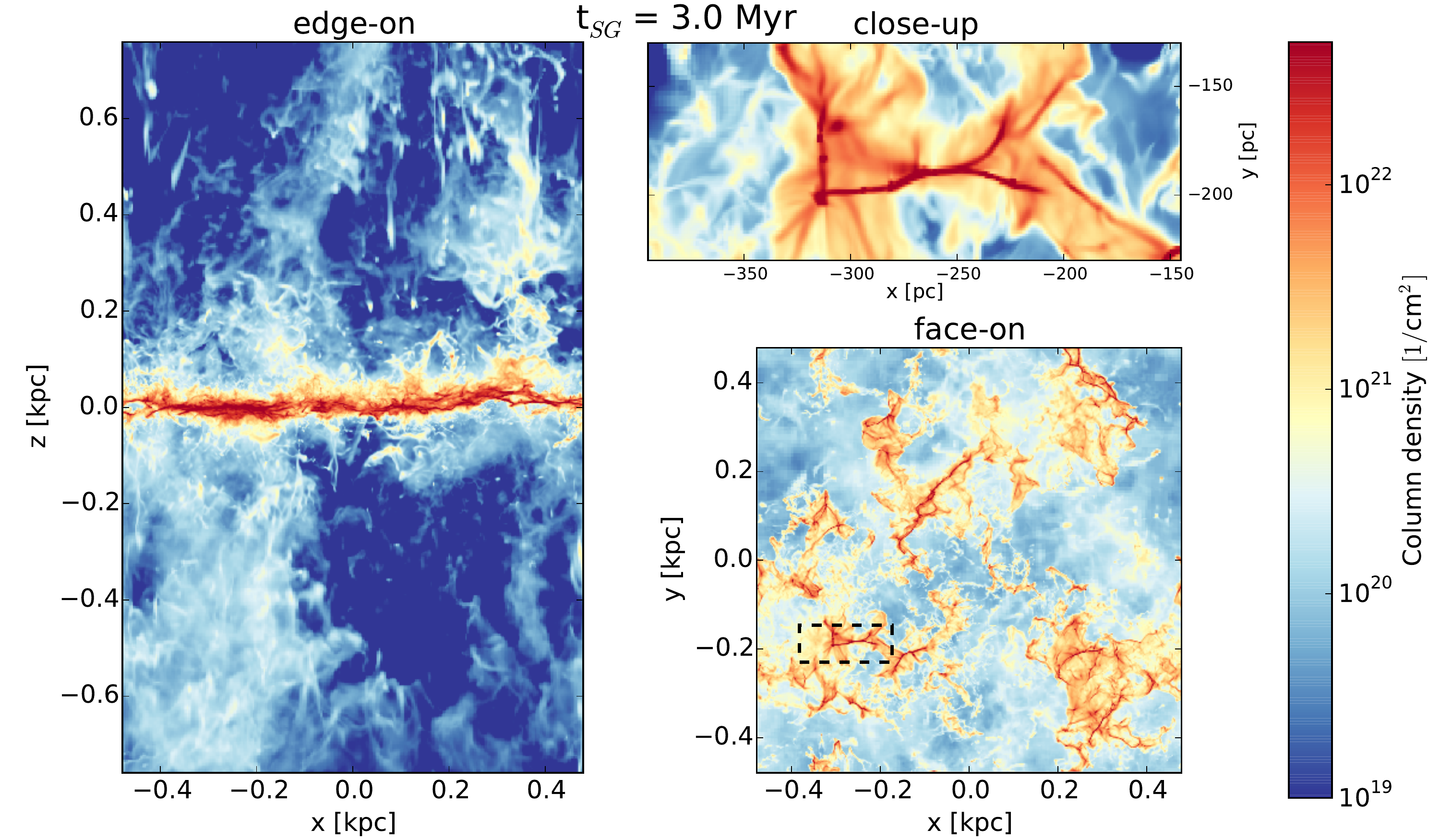} 
\caption{Column density projections at times before and after self-gravity is turned on $t_{SG}= 0$ and~3~Myr.
Each panel shows (left) an edge-on projection of the inner $1 \textrm{~kpc} \times 1.5 \textrm{~kpc}$ of the simulated volume; (bottom right) a face-on projection of the simulated volume, with a $1 \textrm{~kpc}^2$ footprint; and (top right) 
a close-up of the structured, irregular, dense cloud shown with a dashed box in the bottom right panel.
An animation of the self-gravitating evolution of the simulation during the time $t_{SG}=0$--6~Myr is available online.} 
\end{figure*}

\begin{figure*}
\centering 
\label{fig:close-up_slices}
\includegraphics[width=1\textwidth]{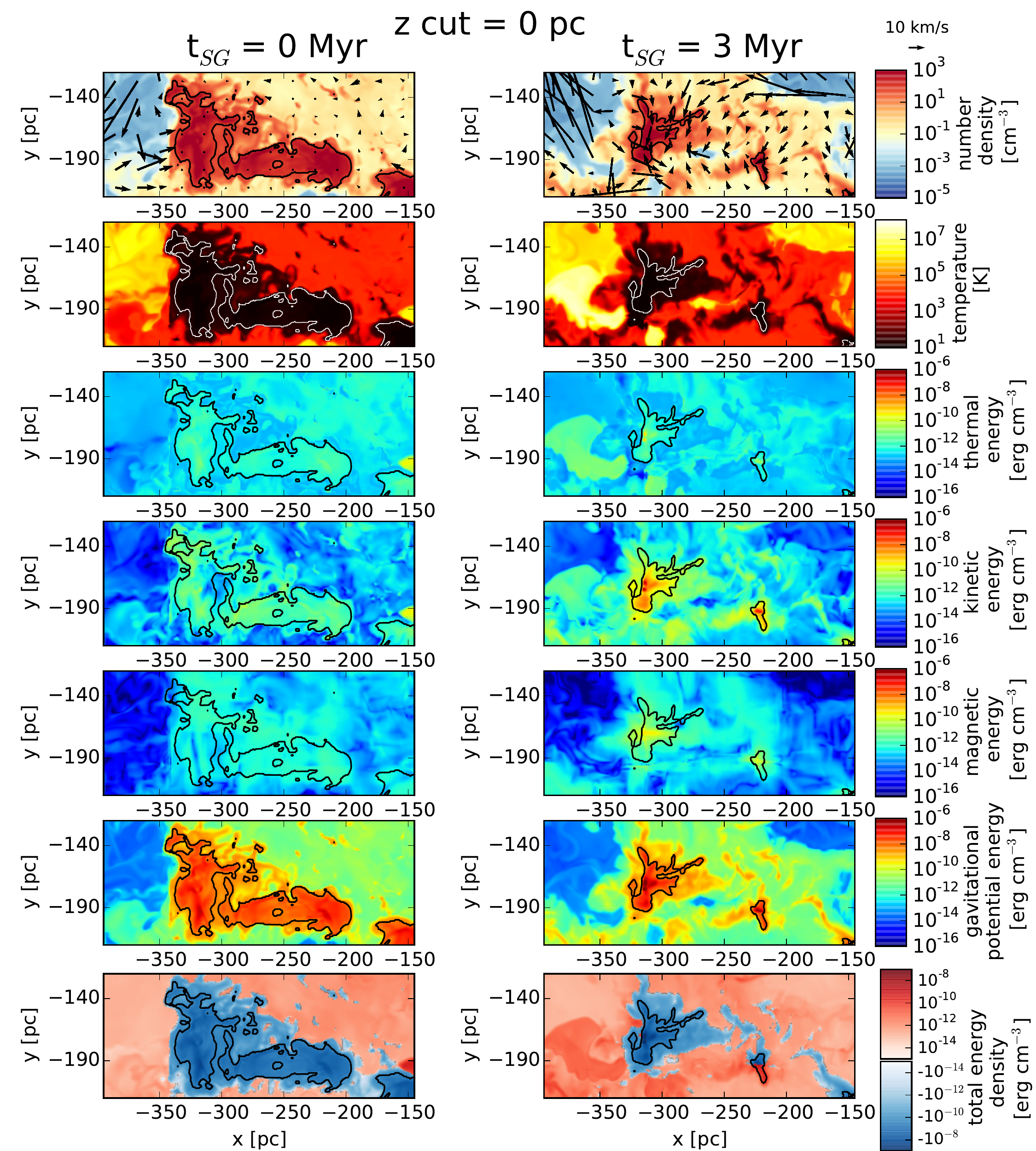} 
\caption{Slice plots of number density, temperature, thermal energy density, kinetic energy density, magnetic energy density, gravitational potential energy and the total energy density are shown. For the closeup cloud seen in projection in Figure~\ref{fig:StratBox_composite} at times before (left) $t_{SG} = 0$ and after (right) $t_{SG} = $~3~Myr self-gravity.
The slice lies in the $x$-$y$ plane, at the midplane $z = 0$.
A black contour denotes the cloud boundary in each slice.
Velocity vectors in the $x$-$y$ plane are included to the number density slice in the first row.
An animation slicing through this region every parsec for altitudes $|z| \leq 50$~ pc is available online. Note that the projection plot of Figure~\ref{fig:StratBox_composite} captures features at multiple altitudes that do not all appear in any single slice.} 
\end{figure*}

\subsection{Overview}

We use the stratified box simulation at time $t = 230 \text{ Myr}$ as a turbulent initial condition for our self-gravitating model, starting with a maximum resolution of 0.95~pc.
At this point, 7,515 SNe have exploded, so the idealized initial conditions of the simulation have long since been erased.
The multiphase ISM has reached a dynamical steady state, where the mass and volume filling fractions of the different ISM phases remain constant in time \citep{Hill2012VerticalMedium}. After we turn on self-gravity, we evolve the simulation for another $6$~Myr.
We stop the simulation at that time because we expect that stellar feedback, particularly ionizing radiation, from the stars formed in the gravitationally collapsing regions will dominate the subsequent evolution \citep[e.g.][]{Dale2012IonizingClusters,Walch2012DispersalRadiation}.

Figure \ref{fig:StratBox_composite} shows our simulation at the moment when self-gravity is turned on,  $t_{SG} = 0$, and at $t_{SG} = $~3~Myr. 
At $t_{SG} = 0$, the gas morphology shows strong stratification, with a dense midplane, and a complex atmosphere. 
Above the midplane, outflows produced by SN explosions and inflows arising from cooling and disk gravity drive gas circulation in a fountain-like manner \citep{Shapiro1976ConsequencesMedium, Bregman1980TheClouds}. The face-on and close-up views show the multiphase structure of the ISM with dense, irregularly shaped clouds that contain most of the mass lying near the midplane.

This cloud population shows a generally filamentary structure, but with filaments that on close examination are broad and diffuse. 
Once self-gravity becomes active, these clouds begin to collapse inward along their shortest dimensions to form far denser and thinner structures.
As these filaments continue to collapse, they begin fragmenting along their lengths, forming dense clumps.
Altogether, we find a complex network of coherent filaments that twist and bend and intersect each other, reaching lengths up to $\sim 200 $~pc.

Figure \ref{fig:close-up_slices} shows slices parallel to the midplane through the cloud shown in detail in the previous figure at $t_{SG} = 0$ and ~3~Myr.
The number density shows a steep gradient at the cloud surface, where a difference of about two to three orders of magnitude occurs between the cloud and the diffuse ISM \citep{Banerjee09ClumpFormation}. 
At later times, this gradient becomes steeper as the cloud collapses.
A similarly sharp gradient is present in the temperature, where a transition between the cold ($\sim$30~K) cloud and the warm ($\sim 10^{4}$ K) ISM occurs.

It is important to remember that we have neglected two important cloud destruction processes that will limit their masses and sizes: galactic rotation and stellar feedback.
Galactic rotation induces shear that will stretch the filaments and tear apart the largest clouds.
Star formation and the resulting stellar feedback will likely destroy the parent clouds on a timescale comparable to the crossing time. 
Because of the lack of either of these effects in our simulations, the clouds live far longer than a crossing time  during the non-self-gravitating evolution, allowing clouds to accumulate mass and grow substantially larger than would be possible otherwise \citep[see discussion in][]{Girichidis2016SILCC}.

\subsection{Cloud Population}

\begin{figure} 
\label{fig:tff_histogram}
\includegraphics[width=0.5\textwidth]{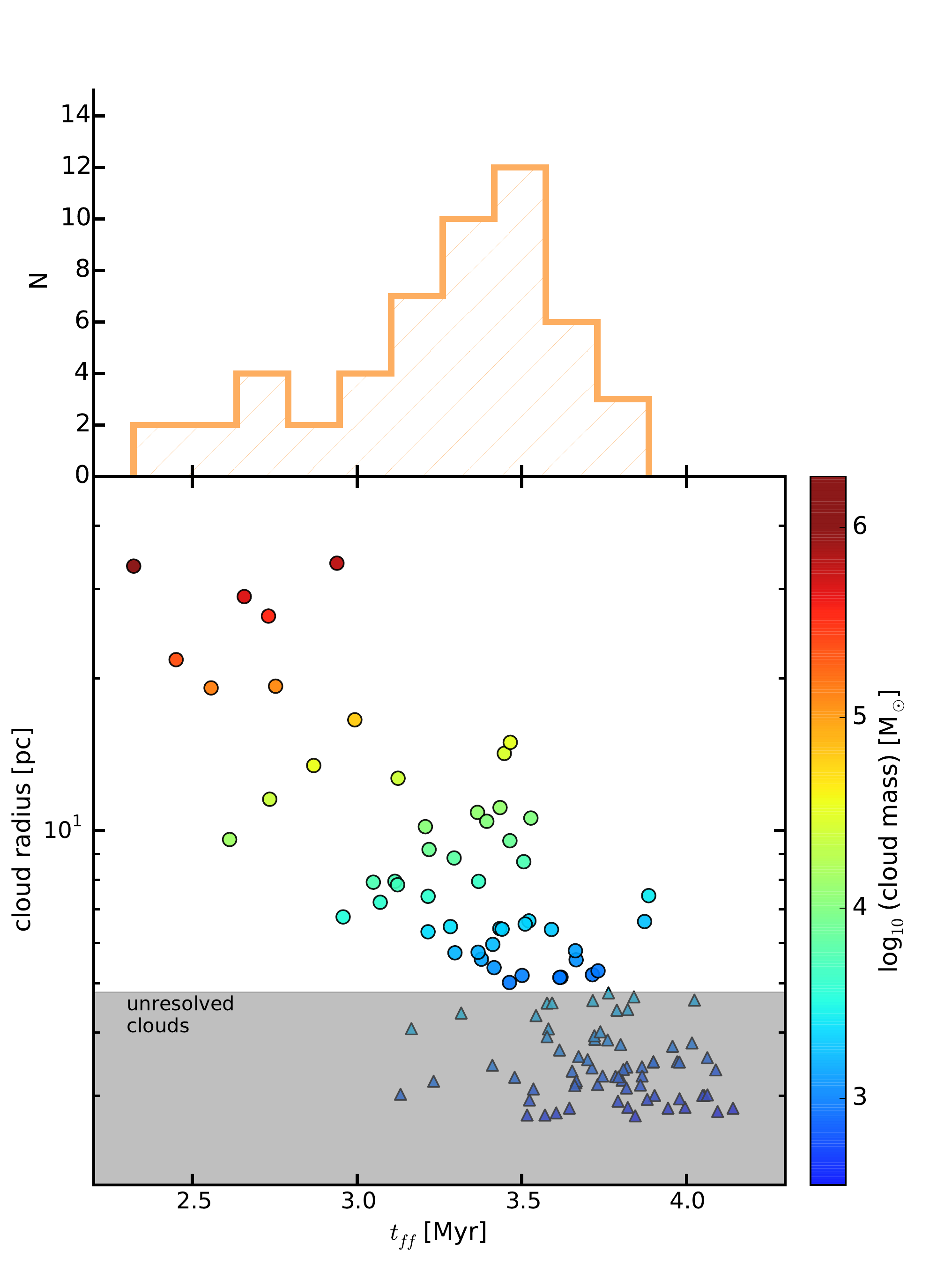}
\caption{Radius $R$, free fall time $\tff$, and mass, $M_{\rm{low}}$, (shown in color) of the simulated cloud population extracted at the time when self-gravity is turned on $t_{SG} = 0$. 
The shaded region in the radius--$\tff $ plot lies below the resolution limit for clouds in the simulation, $R < 5 \Delta x$.
On top a histogram of $\tff$ including only the resolved clouds.} 
\end{figure}

We now follow the formation, fragmentation, and collapse of dense structures formed in our simulations, extracting approximately 40 resolved clouds and 2--4 clumps at each snapshot. 
We track the evolution of these structures from one snapshot to the next using tracer particles. 
We compute the mass $M_{\rho}$, radius $R_{\rho}$, and velocity dispersion $\sigma_{\rho, tot}$, of the clouds and clumps at each time in the evolution.
The initial population is first extracted at the moment self-gravity is turned on, $t_{SG}= 0$, corresponding to a global evolutionary time of $t=230$~Myr.
Figure \ref{fig:tff_histogram} shows the basic properties of our initial cloud population.
The simulated clouds span a wide range in radii $4.8 $~pc $< R < 40$~pc, masses $1.3 \times 10^{3} $~M$_{\odot} < M <2 \times 10^{6} $~M$_{\odot}$, and mean densities $10^{2} $~cm$^{-3}< n < 3 \times 10^{3} $~cm$^{-3}$, corresponding to a range of free fall times of $2 $~Myr$ < \tff  < 4 $~Myr.
The simulated cloud mass function is consistent for different resolutions $\Delta x = 0.47, 0.95, 1.9$, and 3.8~pc, and for different global evolutionary times, $t=100, 150$, and 300~Myr. 
Most of our clouds are located at distances $|z| <$~50~pc from the midplane, in rough agreement with the observed scale height of the molecular gas in the Galaxy \citep{Clemens1988TheQuadrant}.
We focus our presentation on results from our simulations with $\Delta x = 0.95$~pc resolution. 
We also show resolution studies that reveal numerical effects on the measurement of the velocity dispersion in our simulations.

\subsection{Virial Balance Evolution}

The evolution of molecular clouds is determined by the interplay between thermal energy, turbulence, magnetic fields and gas self-gravity.
Rows 3~-~7 in Figure \ref{fig:close-up_slices} show slices of the different energy densities that govern the dynamics of the cloud.
Snapshots at two evolutionary times are shown, left, at the moment self-gravity is turned on, $t_{\rm{SG}}= 0$, and right, at 3~Myr after self-gravity has been active.
The thermal energy is roughly uniform throughout the cloud and its environment. 
The highest variation observed in this slice corresponds to an expanding SN remnant outside the cloud reaching thermal energies three to four orders of magnitude higher than its surroundings. 
The overall contribution of the thermal energy compared to the other components is very low.
Although the turbulent velocities inside the cloud are slower than expected (see discussion in section \ref{sec:sigma-R}), the kinetic energy inside the cloud exceeds that of the background because of the high densities in the cloud. 
A significant increase in the kinetic energy is observed at later times as gas falls towards local centers of gravitational collapse throughout the cloud.
The magnetic energy shows little variation between the cloud interior and its surroundings at $t_{\rm{SG}}=0$.
At later times a significant increase in the magnetic energy is observed as the cloud contracts and the magnetic field is compressed, but the magnetic energy remains subdominant.
The gravitational potential energy dominates the overall energy budget of the cloud everywhere, most significant in regions where the density is highest.
As the cloud contracts, some of the gravitational potential energy is converted into kinetic and magnetic energy, but the gravitational potential energy also deepens at the centers of collapse.
The bottom row of Figure \ref{fig:close-up_slices} shows the sum of the volume energy densities contributing to the cloud energetics, neglecting the surface terms \citep{McKee1992} something that we will examine in the future.
It is clear that for both snapshots the cloud and its environment is dominated by the gravitational potential energy.
This leads to gravitational collapse of the cloud.

\begin{figure} 
\centering 
	\centering 
	\label{fig:alpha-virial}
	\includegraphics[width=0.48\textwidth]{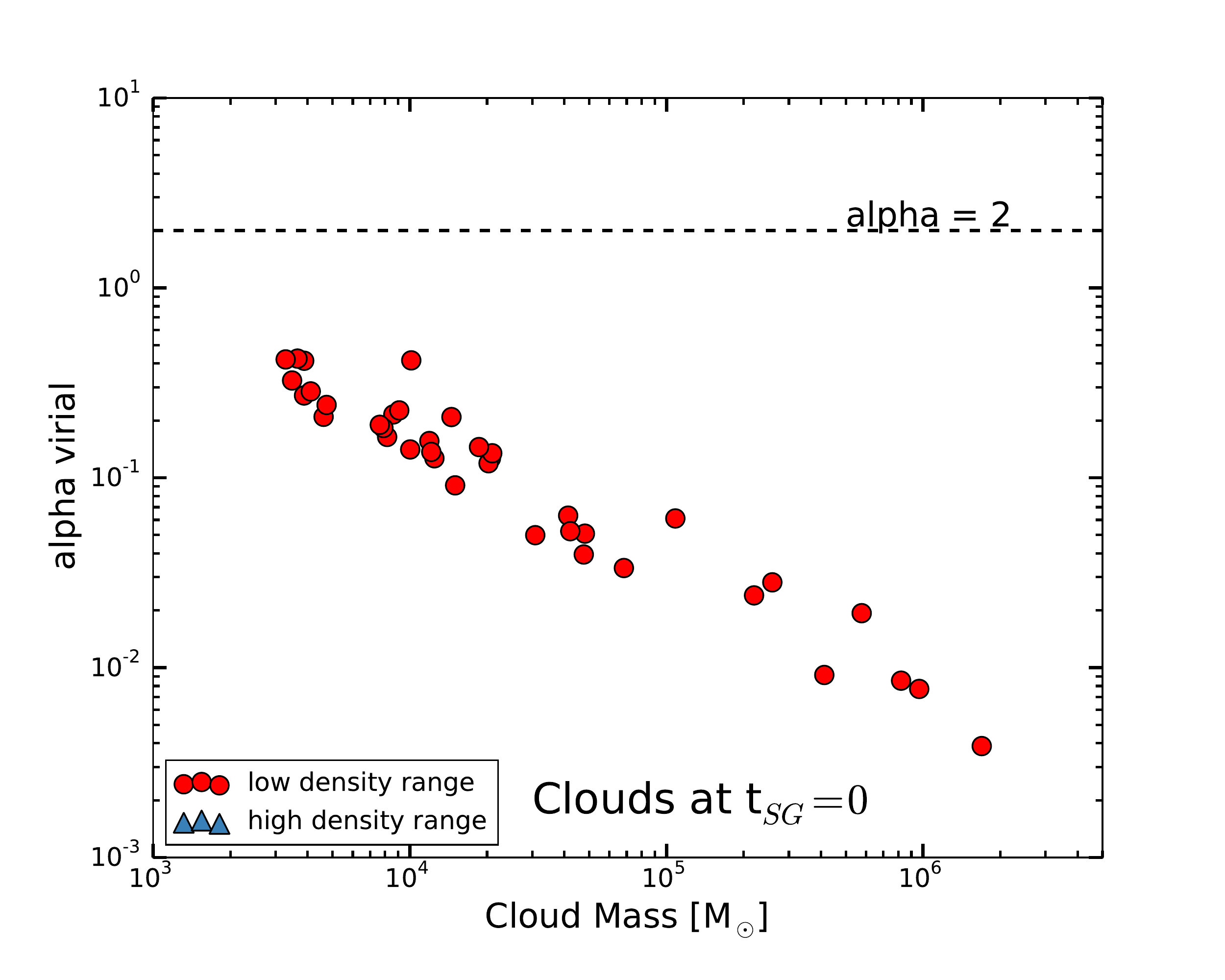} \\
	\includegraphics[width=0.48\textwidth]{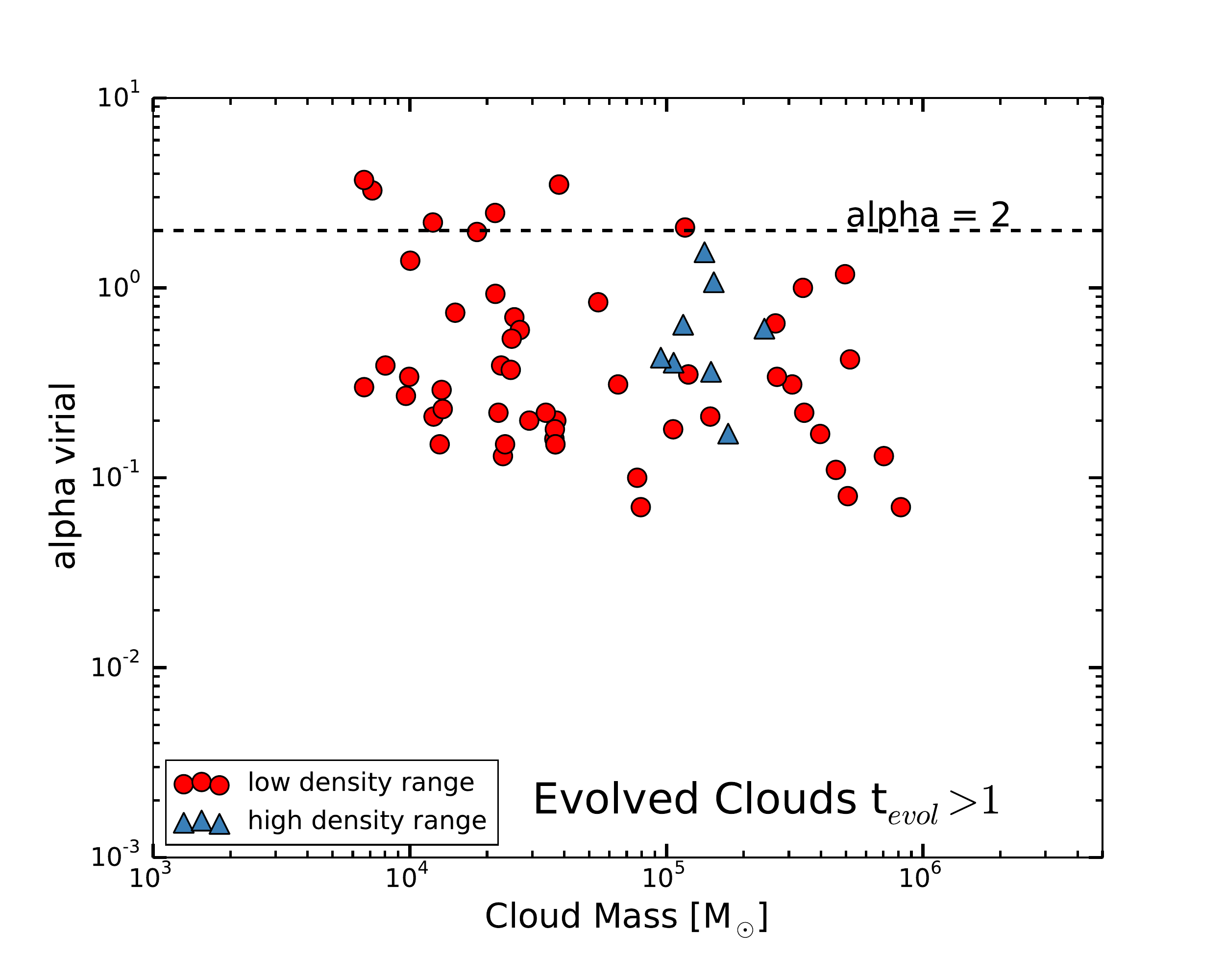}
\caption{Virial parameter vs mass plot for the (top) cloud population extracted at $t_{\rm{SG}}=0$ and (bottom) evolved simulated cloud population $t_{\rm{evol}}>1$.
(both panels) Black dashed line corresponds to $\alpha_{virial} = 2$ equivalent to a system in virial equilibrium $|E_{k}|=2|E_{g}|$.} 
\end{figure}

We argue that the clouds are in a constant state of gravitational collapse.
In order to further explore this idea we examine the behavior of the simplified virial parameter often used in studies of molecular cloud dynamics \citep{Bertoldi92VirialParameter,BallesterosParedes06VirialParameter, Kauffmann13LowVirial}, given by 
\begin{eqnarray}
	\alpha_{vir} = \frac{5 \sigma_{\rho, tot}^2 R_{\rho}}{  G M_{\rho}}, 
\end{eqnarray}
for spherically symmetric clouds, where $\sigma_{\rho, tot}$ is the total velocity dispersion for a given density range, $R_{\rho}$ the radius and M$_{\rho}$ is the mass within that same density range.
Figure \ref{fig:alpha-virial} shows the virial parameter for the simulated cloud population, revealed by our low density tracer, at $t_{\rm{SG}}=0$, and for the evolved cloud population, $t_{\rm{evol}}>1$ (see Equation \ref{eq:tevol}).
Nearly all our clouds are bound and unstable, with $\alpha_{vir} < 2$, particularly before self-gravity has begun to affect the cloud dynamics. 
At later times, when self-gravity has driven fast, chaotic motions, virial parameters are much higher, though only a few of them reach the marginally stable regime, $1 \leq \alpha_{vir} \leq 2$.

\subsection{Evolution of the Velocity Dispersion-Radius Relation}
\label{sec:sigma-R}

\begin{figure*} 
\label{fig:sigma-L-evolution}
\includegraphics[width=1.0\textwidth]{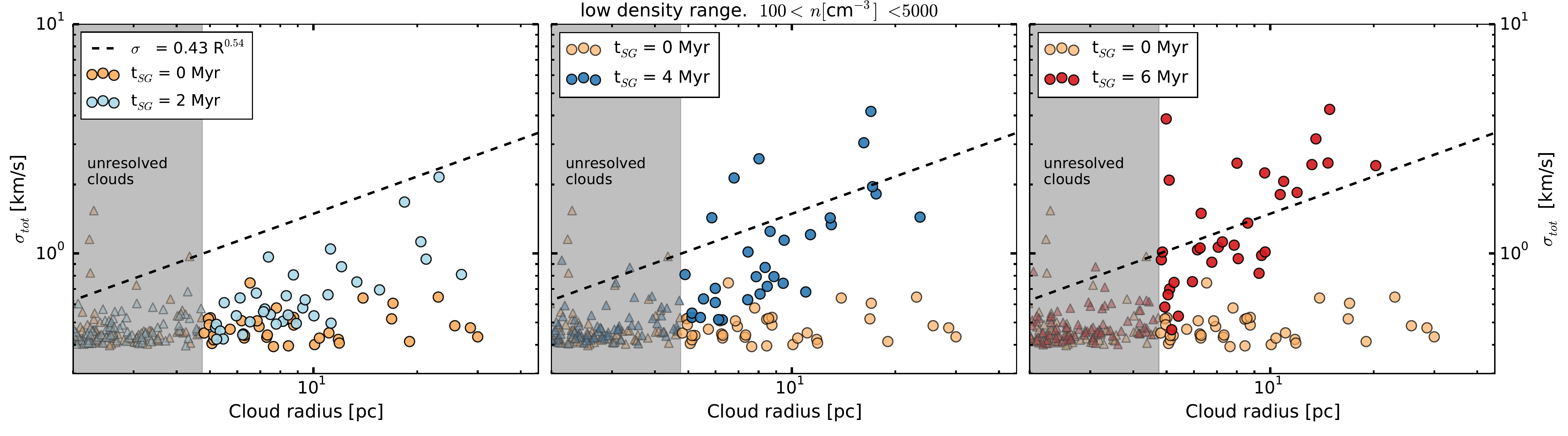}
\includegraphics[width=1.0\textwidth]{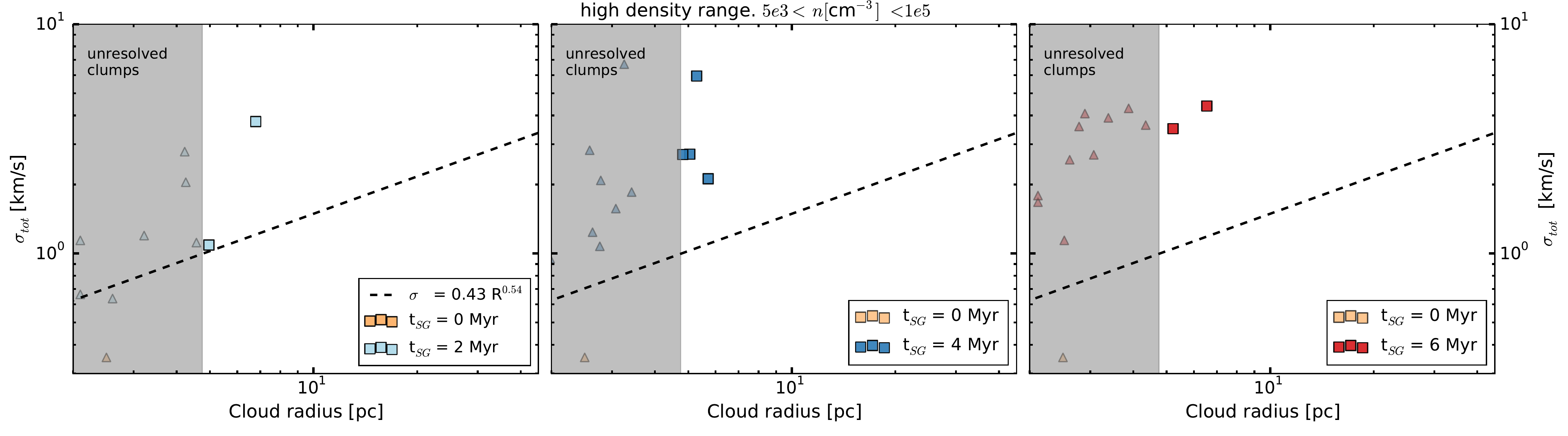}

\caption{Velocity dispersion-radius relation for different density ranges, at different times after self-gravity is turned on in the simulations.
For all plots: dashed line corresponds to our best fit to the GRS data of the velocity dispersion-radius relation $\sigma_{tot} =0.43 (R/\mbox{1 pc})^{\,0.54}$~km~s$^{-1}$, the shaded gray region distinguishing unresolved from resolved objects.
Columns show (left to right) evolutionary time $t_{SG} = 2, 4,$ and~6~Myr, while rows show different density tracers (top to bottom)  $100$~cm$^{-3} <  n_{\rm{low}}   < 5000$~cm$^{-3}$ and $5\times10^{3}$~cm$^{-3} < n_{\rm{int}} <1\times10^{5}$~cm$^{-3}$. All plots contain the objects extracted at that same density range prior to the action of self-gravity  ($t_{SG} = 0$) in orange.} 
\end{figure*} 

Figure \ref{fig:sigma-L-evolution} shows the evolution of the total velocity dispersion, $\sigma_{tot}$, vs. the cloud radius, $R$, for the different density ranges at successive times, $t_{SG}$, after self-gravity is turned on. 
At $t_{SG} = 0 $ the clouds captured by the low density tracers disagree with Larson's relations in both slope and normalization.
At this time, clouds have very low velocity dispersions, $0.35 $~km~s$^{-1} < \sigma < 0.6 $~km~s$^{-1}$ that show no correlation with their radius.
There is a complete absence of dense and compact structures traced by the high density tracer at $t_{SG}= 0$.

As self-gravity acts, clouds quickly react to this new force, with radius shrinking and internal motions increasing.
Figure \ref{fig:tff_histogram} showed that larger clouds are more massive and have shorter free fall times, because they tend to have larger average densities.
Consequently these clouds react most strongly to self-gravity, which increases their velocity dispersions. 
As the simulation including self-gravity evolves, the clouds contract and the gas within them begins flowing towards higher and higher densities while at the same time the clouds continue growing in mass through accretion of ambient material. 
This processes drives high velocity dispersions that after several megayears begin to show a correlation with the cloud size in agreement with Larson's relations.
Clumps, captured by the high density tracer, increase their velocity dispersions, overshooting Larson's relations, in agreement with observations of high density tracers \citep{Caselli1995TheCores, Plume1997DenseMasers, Gibson2009MOLECULARCONDITIONS}.
They collapse faster than their surrounding envelopes, suggesting a hierarchical state of collapse \citep{Elmegreen2007}. 
By $t_{SG} = 6 \text{ Myr}$, clouds and clumps have significantly modified their internal velocity dispersions. 
The structures captured by the low density range, show a velocity dispersion--radius relation similar to Larson's fit (see Section \ref{sec:observations} for quantitative discussion).

\subsubsection{Resolution Study}
\label{sec:Resolution Study}

The results presented in Figure~\ref{fig:sigma-L-evolution} suggest that SN-driven turbulence in the diffuse ISM {\em can not} drive fast turbulent motions in dense clouds.
However many of the clouds presented here are only resolved by 10--20 cells in diameter (a few thousand cells in volume). 
In order to explore the effects of numerical resolution in our results, we run a series of resolution tests.
Starting at $t=230 \, \rm{Myr}$, we run our simulations forward {\em without} self-gravity for $10 \, \rm{Myr}$ at resolutions, $\Delta x = 0.47, 0.95$, and 1.9~pc.

\begin{figure} 
\centering 
\label{fig:sigma-L-Resolution}
\includegraphics[width=0.5\textwidth]{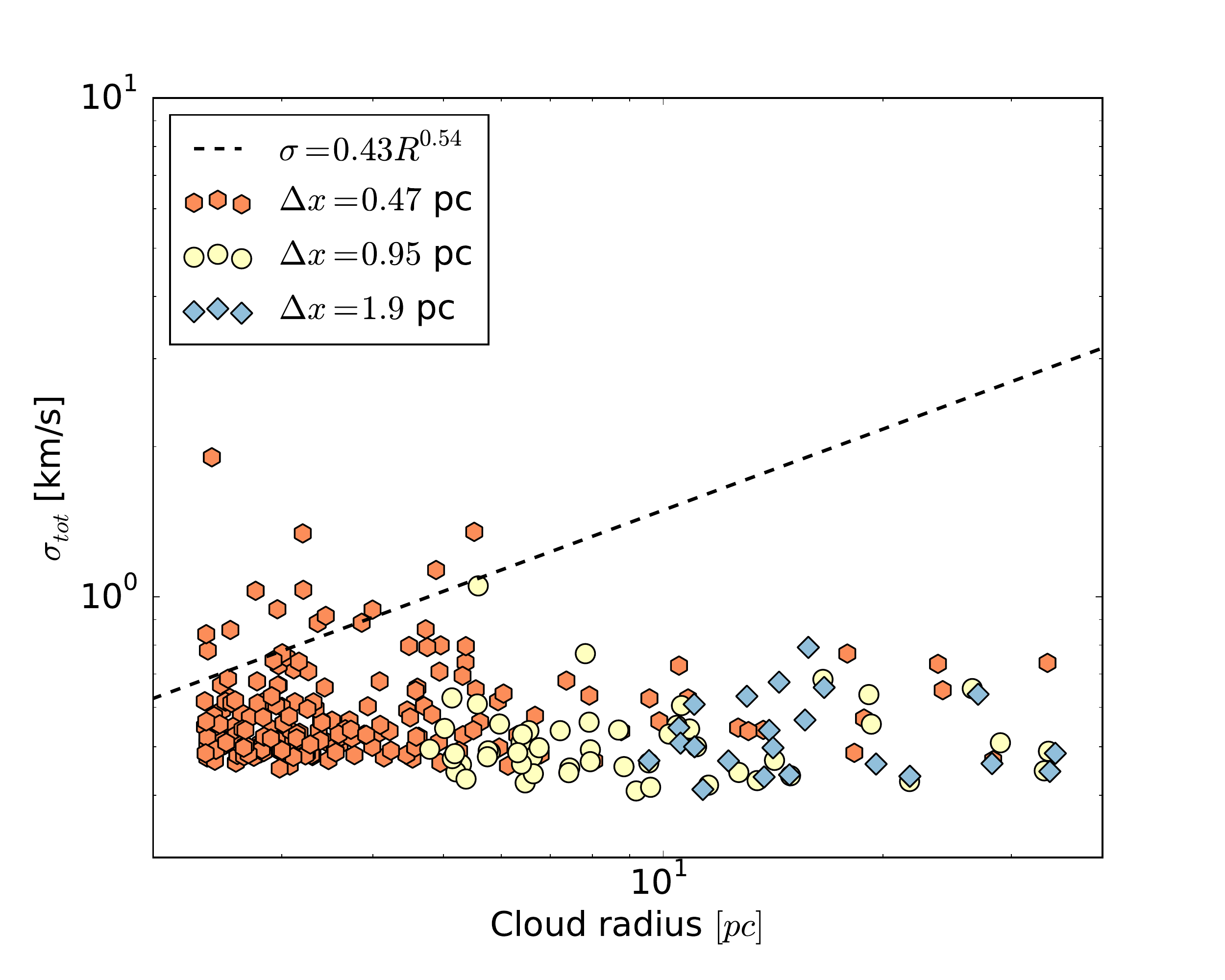}
\caption{Relation between velocity dispersion and radius for the cloud population absent self-gravity at resolutions $\Delta x = 0.47$~pc (orange hexagons), 0.95~pc (yellow circles) and 1.9~pc (blue diamonds). A minimum resolution threshold for the cloud radius of $R \geq 5 \Delta x$ is imposed on all three populations. The dashed line corresponds to our fit to the velocity dispersion-radius relation in the GRS data $\sigma = 0.43 R^{0.54}$~km~s$^{-1}$. } 
\end{figure}

We extract and analyze a cloud population at the final snapshot of each of these simulations.
Figure \ref{fig:sigma-L-Resolution} shows the $\sigma_{tot}-R$ relation for this cloud population at each resolution.
Within the three simulations, the clouds identified have radii in the range $2.5 \text{ pc} < R < 60 \, \text{ pc}$.
At all three resolutions, clouds have low velocity dispersions uncorrelated with radius. 

As discussed by \citet{Banerjee09ClumpFormation}, as clouds grow, they are unresolved during the initial stages of their formation, but later become resolved as they reach sufficiently large sizes.
The high resolution simulation, with $\Delta x = 0.47$~pc, resolves small objects at early stages of cloud formation and evolution, so we see more variation of the velocity dispersion in small clouds for the $\Delta x = 0.47$~pc resolution simulations, compared to the $0.95$~pc and $1.9$~pc resolution simulations shown in Figure \ref{fig:sigma-L-Resolution}.

Because of the absence of internal feedback that might destroy the clouds in our simulations, they live long lives.
Long lived clouds have enough time for their internal turbulence to decay \citep{MacLow1998KineticClouds,Stone1998DissipationTurbulence}.
These structures then maintain low internal velocity dispersions while ambient SN-driven turbulence cannot drive strong turbulent motions inside the dense clouds.

\subsection{Quantifying Cloud Evolution}

We want to extract a cloud population that can be directly compared with observations, but we believe that the quiescent clouds at $t_{SG} = 0$ are unrealistic because of their long lives and low velocity dispersions. 
The gravitationally evolved clouds, on the other hand, appear more physical.  
Therefore, we wish to distinguish the evolving clouds and compare only them to the observations.  
To do this, we follow the evolution of individual clouds through time and quantify their evolution, in order to classify them as quiescent or evolved.
For this, we introduce the normalized evolutionary timescale, the ratio of the time self-gravity has been active to the cloud's initial free fall time
\begin{eqnarray}
 t_{\textrm{evol}} = t_{SG} \,/\, t_{ff} (t_{SG} = 0).
 \label{eq:tevol}
\end{eqnarray}

Most of our clouds are indeed present when self-gravity is turned on, and have initial properties taken at that time.  
However we also identify a number of clouds formed during the self-gravitating period of the simulation. 
These clouds are extracted separately.
Their initial properties are taken at the time they were first identified as resolved clouds.

\begin{figure*} 
\centering 
\label{fig:sigma-L-evolution_tff}
\includegraphics[width=1.0\textwidth]{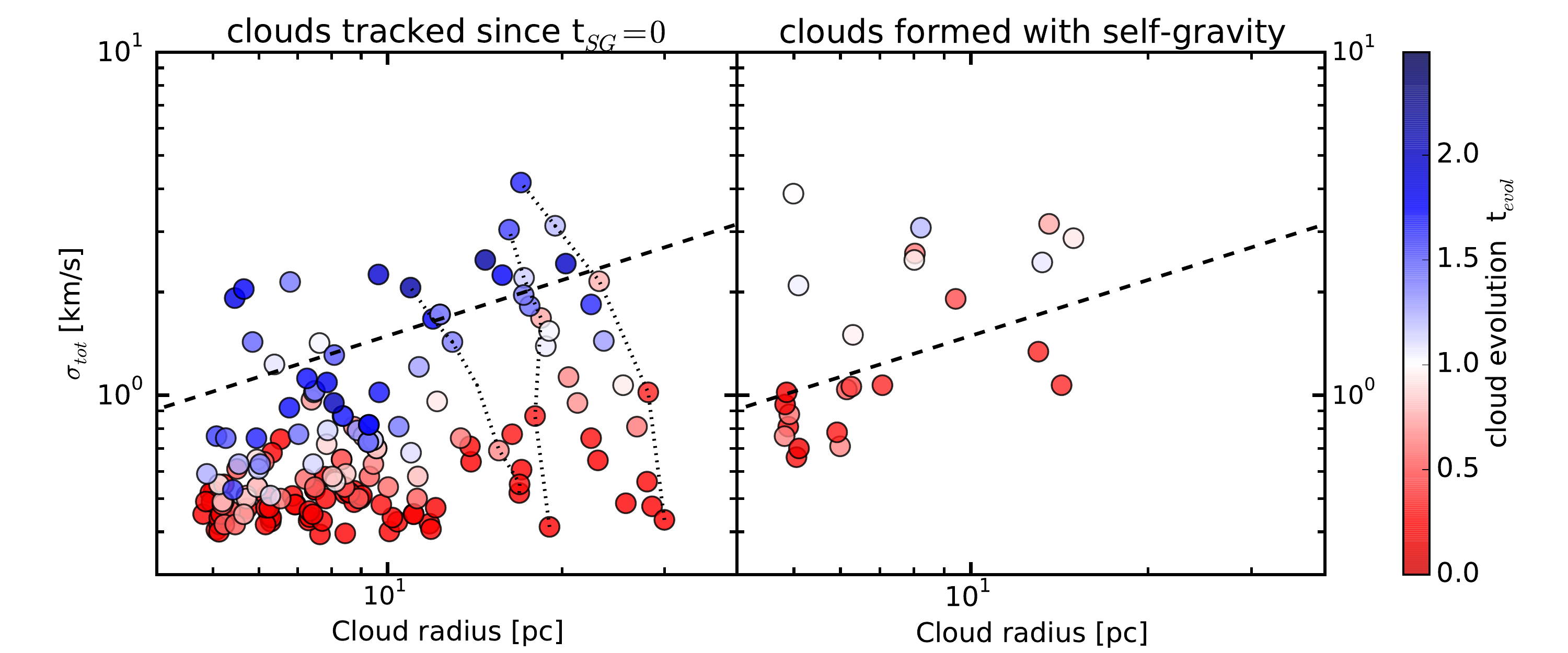}
\caption{Velocity dispersion as a function of cloud radius for clouds tracked during the self-gravitating evolution of the simulation, (left) clouds present at $t_{SG} = 0$, and (right) clouds formed after self-gravity was turned on, $t_{SG} > 0$. 
Clouds are identified in snapshots at $t_{SG} = 0, 1, 2, 3, 4, 5$, and~6~Myr, and are colored by the evolutionary timescale $t_{evol} = t_{SG} / \tff $. For three cases, dotted lines track the evolution of specific clouds through all snapshots. The dashed line corresponds to our fit to the GRS data $\sigma = 0.43 R^{0.54}$~km~s$^{-1}$.} 
\end{figure*}

We combine all the clouds identified at times $t_{SG}=0, 1, 2, 3, 4, 5$, and~6 Myr in order to have a mixed population of clouds at different evolutionary stages.
Figure \ref{fig:sigma-L-evolution_tff} shows the compilation of these clouds traced by the low density range tracer. 
The evolutionary timescale $t_{evol}$ allows us to differentiate clouds at different stages of their evolution. 
A clear distinction can be seen between clouds that have evolved to $t_{evol} > 1$ and those that have not yet reached that point.
Clouds that have $t_{evol} > 1$  show higher velocity dispersions and lie close to the expected velocity dispersion-radius relation.
Clouds with $t_{evol} < 1$, on the other hand, show low velocity dispersions remaining from their quiescent evolution during the non-self-gravitating period.

From Figure~\ref{fig:sigma-L-evolution_tff}, it is clear that gravity can increase the internal velocity dispersion of a cloud in a free fall time.
The right panel of Figure~\ref{fig:sigma-L-evolution_tff} shows clouds formed during the self-gravitating period of the simulations.
These clouds still preserve some turbulence left over from their formation and show internal velocity dispersions systematically higher than those of the long-lived, quiescent clouds formed during the non-self-gravitating evolution of the simulation.
We use only the population of clouds with $\rm{t}_{evol} \geq 1$ and the clouds first formed during the self-gravitating period of the simulations to compare with the observations.

\subsection{Comparison With Observations}
\label{sec:observations}

The Boston University FCRAO Galactic Ring Survey (GRS) is a molecular line survey of the inner Galaxy. 
It offers excellent sensitivity ($<0.4$~K), high spectral resolution (0.2~km~s$^{-1}$), angular resolution of 46" and sampling of 22"  
\citep{Sanders1986Massachusetts-StonyQuadrant,Clemens1986Massachusetts--StonyQuadrant,Jackson2006TheSurvey,Roman-Duval2010PHYSICALSURVEY}. 
This survey uses $^{13}$CO$(1-0)$, which is more suitable for studying dynamics than the commonly used $^{12}$CO in previous studies of the Larson's relations \citep{Larson1981,Solomon1987MassClouds}.
This is because $^{13}$CO is some 30--70 times less abundant than $^{12}$CO \citep{Langer1990C-12/C-13Clouds}, so it remains optically thin on parsec scales. 
Therefore $^{13}$CO observations have a higher dynamic range of gas column densities than $^{12}$CO.
We use a subset of the GRS survey here, similar to the data used by \citet{Heyer2009RE-EXAMININGCLOUDS}, corresponding to the same clouds observed by \citet{Solomon1987MassClouds} in their examination of the Larson relations.
We perform a Bayesian parameter estimation of the velocity dispersion-size relation to this data. We obtain a posterior distribution for the intercepts with a $2\sigma$ high density interval (HDI) of [0.20, 0.82] and a distribution of slopes with a $2\sigma$-HDI [0.31, 0.81]. 
We take the posterior median slope, 0.54, and median intercept, 0.43, as the canonical $\sigma-R$ relation in all of our plots.

\begin{figure} 
\centering 
\label{fig:sigma-L-evolution_GRS}
\includegraphics[width=0.48\textwidth]{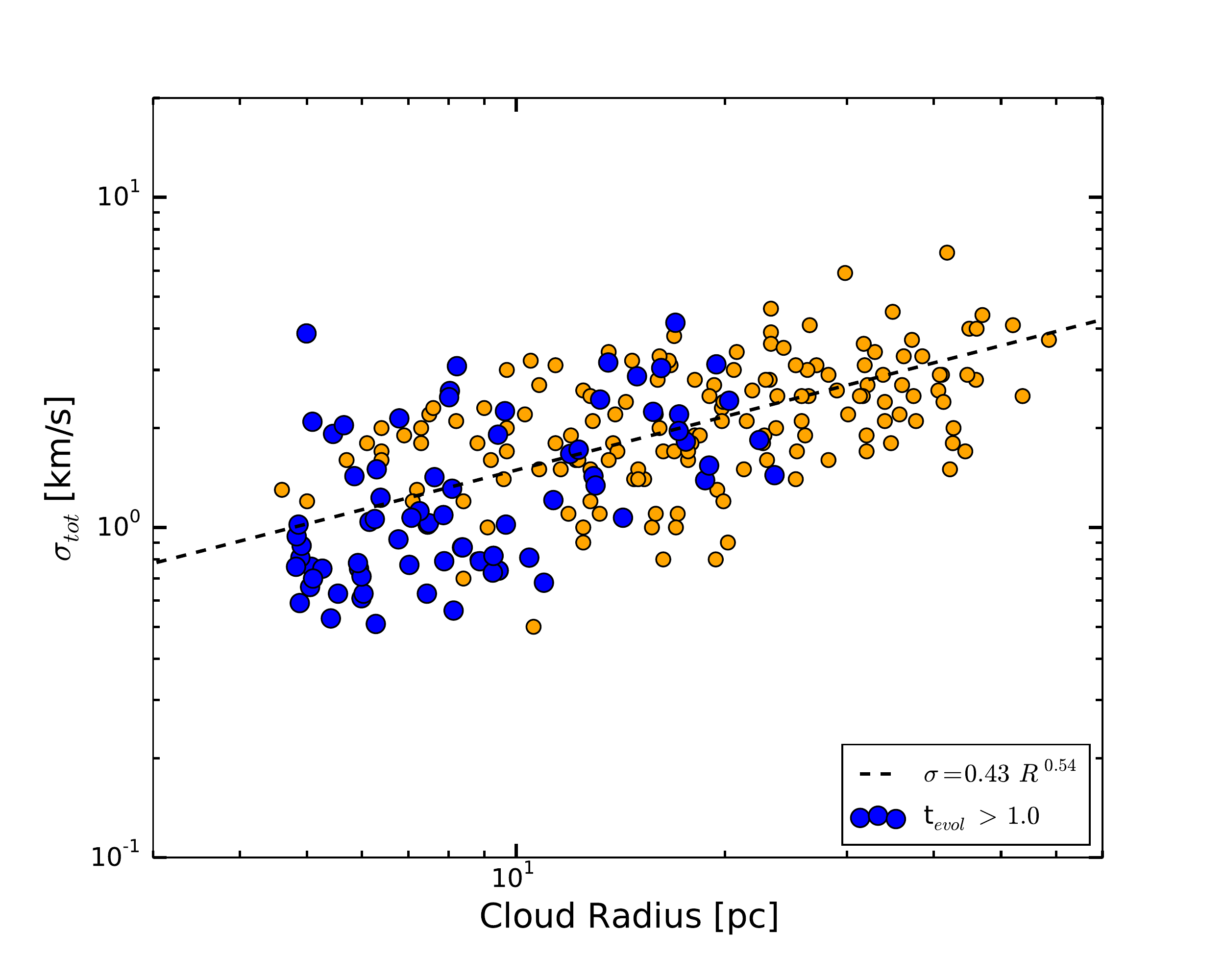}
\caption{Velocity dispersion-radius relation for the subset of the observed GRS cloud catalog used by \citet{Heyer2009RE-EXAMININGCLOUDS} compared to the simulated, resolved, {\em evolved} clouds with $t_{evol} \geq 1$ as well as the clouds formed at $t_{SG} > 0$. The dashed line corresponds to our best fit of the GRS data $\sigma = 0.43 R^{0.54}$~km~s$^{-1}$.} 
\end{figure} 

Figure \ref{fig:sigma-L-evolution_GRS} shows the $\sigma_{tot}-R$ relation for the GRS clouds and the evolved cloud population from the simulations.
The evolved population of simulated clouds have increased their velocity dispersions exhibiting a correlation with the cloud size, now closely resembling the observed GRS cloud population. 
We emphasize that the lack of correlation in the model without self-gravity is at least as important to our understanding of the dominant physics as the correlation seen in the self-gravitating model.
\begin{figure} 
\centering 
\label{fig:sim_clouds_slope}
\includegraphics[width=0.48\textwidth]{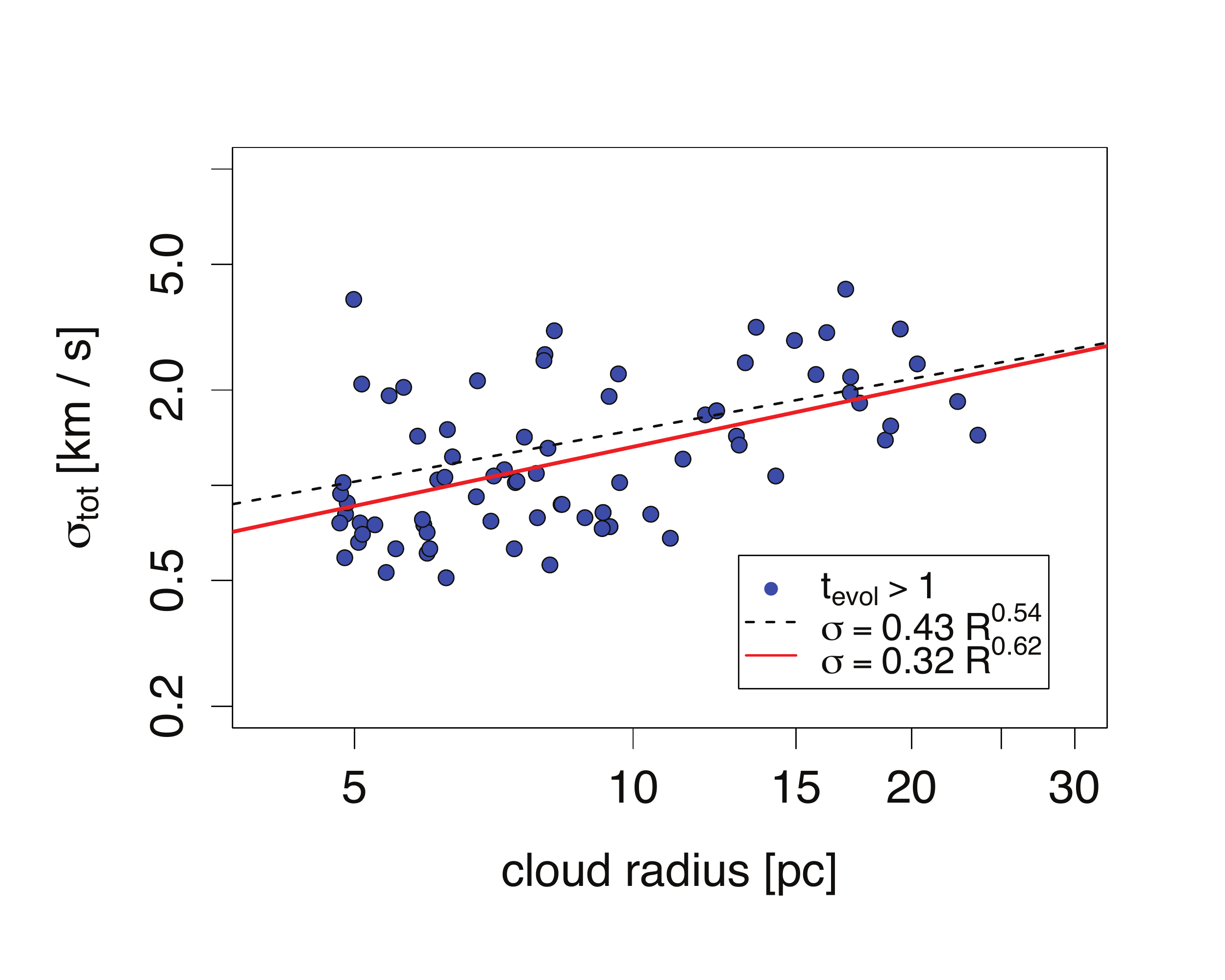}
\caption{Velocity dispersion-radius relation for the simulated cloud population.  
The red line corresponds to the slope and intercept for our cloud population $\sigma_{tot} = (0.32 \mbox{ km s}^{-1}) R_{pc}^{0.62}$.
For reference, the black dashed line is our fit to the GRS data $\sigma_{tot} = (0.43 \mbox{ km s}^{-1}) R_{pc}^{0.54}$. } 
\end{figure} 

Figure \ref{fig:sim_clouds_slope} shows the simulated cloud population along with the power law regression model, $\sigma = K R_{pc}^{\alpha}$, applied to that population. We obtain an intercept and a slope of $K = 0.32 \pm 0.11$ and $\alpha = 0.62 \pm 0.12$ respectively, in close agreement to the parameters estimated for the GRS clouds and those derived from other observations \citep{Solomon1987MassClouds, Falgarone2009IntermittencyScale}.
What is notable in our cloud population is that all of our clouds are collapsing gravitationally.
This means that including self-gravity to the SN driven cloud population, was enough to produce velocity dispersions consistent with the observations, suggesting that it is the clouds' gravitational collapse that drives the observed non-thermal linewidths \citep{Lee2015Time-varyingRate, Burkhart2015}.

We recover a normalization for the simulated cloud population of $\sigma_{tot} / R^{1/2}_{pc} = 0.32 \pm 0.11$, lower than the historical values of $1.1$ reported by L81 or $1.0$ by \citet{Solomon1987MassClouds}, but more consistent to the re-examined values, $0.3$ for molecular clouds in the outer Galaxy \citep{Heyer2001EquilibriumGalaxy}, $0.43$ for our parameter estimation of the GRS data or $0.42$ found in numerical simulations by \citet{Padoan2015OriginTurbulence}.
Sources of uncertainty in our results include underestimation of the velocity dispersion and overestimation of the clouds sizes and surface densities, as the Jeans length in the clouds is marginally resolved.
One should be careful when comparing the normalization of the velocity dispersion-radius relation between simulations and observations, because of four factors that directly affect this quantity: First, it has been shown by \citet{Shetty2010THERELATIONSHIPS} that the effects of projection have an effect on the measured normalization, but not on the slope of the $\sigma_{tot}-R$ relation.
Second, analyses of numerical simulations assume truly optically thin emission for the gas, unless a proper treatment of radiative transfer is applied during post-processing, overestimating the amount of emitting material.  
Third, most simulations do not follow the non-equilibrium chemical evolution of the ISM, and so do not predict the real abundances of the various emitting molecules.
And fourth, as pointed out by \citet{Heyer2009RE-EXAMININGCLOUDS}, the velocity dispersion depends not only on the cloud sizes but also on the clouds' surface density, so variation in the analyzed density range can give rise to variation in the normalization.

\subsection{Variable Column Densities }
\label{sec:sigma-R-Sigma}

\begin{figure*} 
\centering 
\label{fig:Heyer-Plot}
\includegraphics[width=0.495\textwidth]{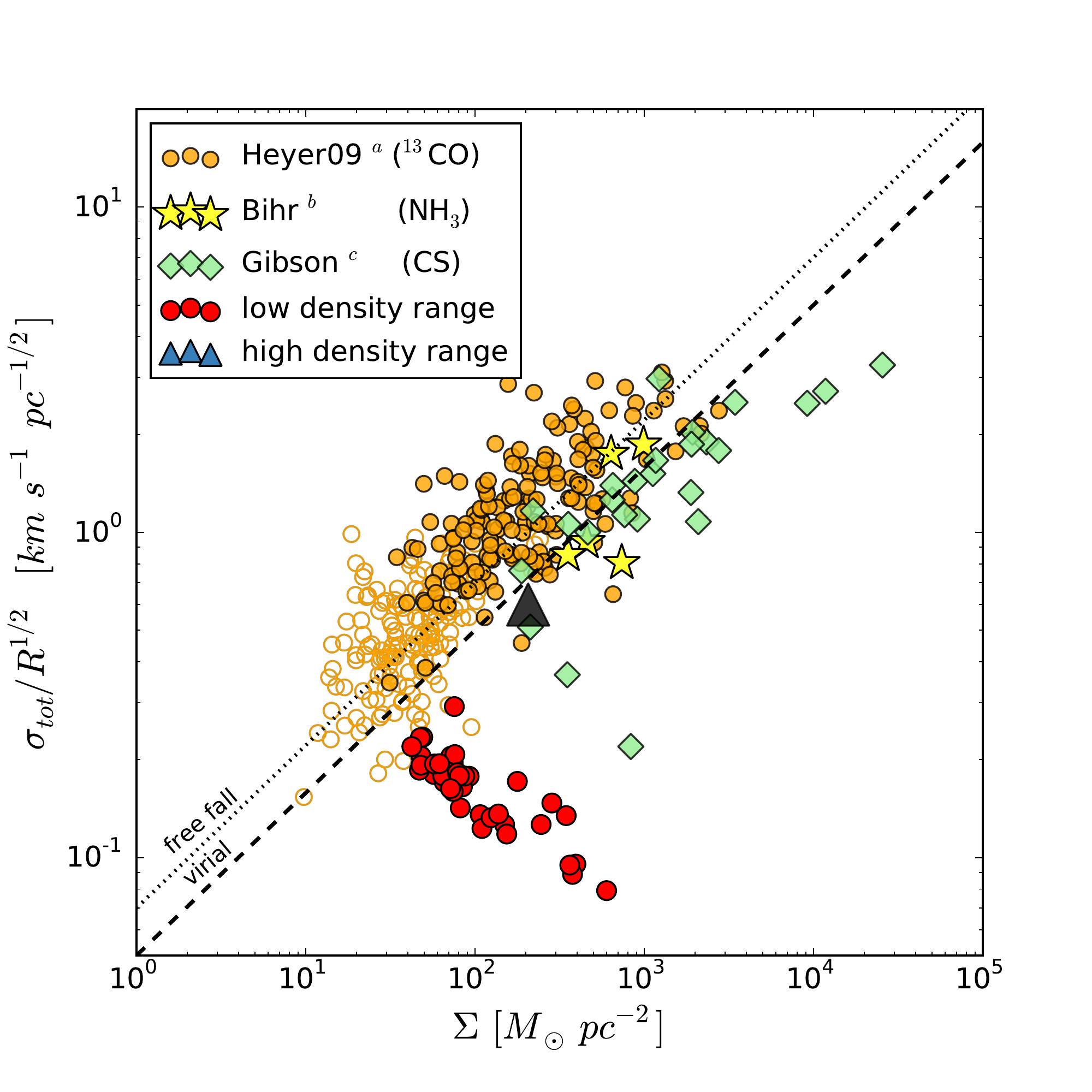} 
\includegraphics[width=0.495\textwidth]{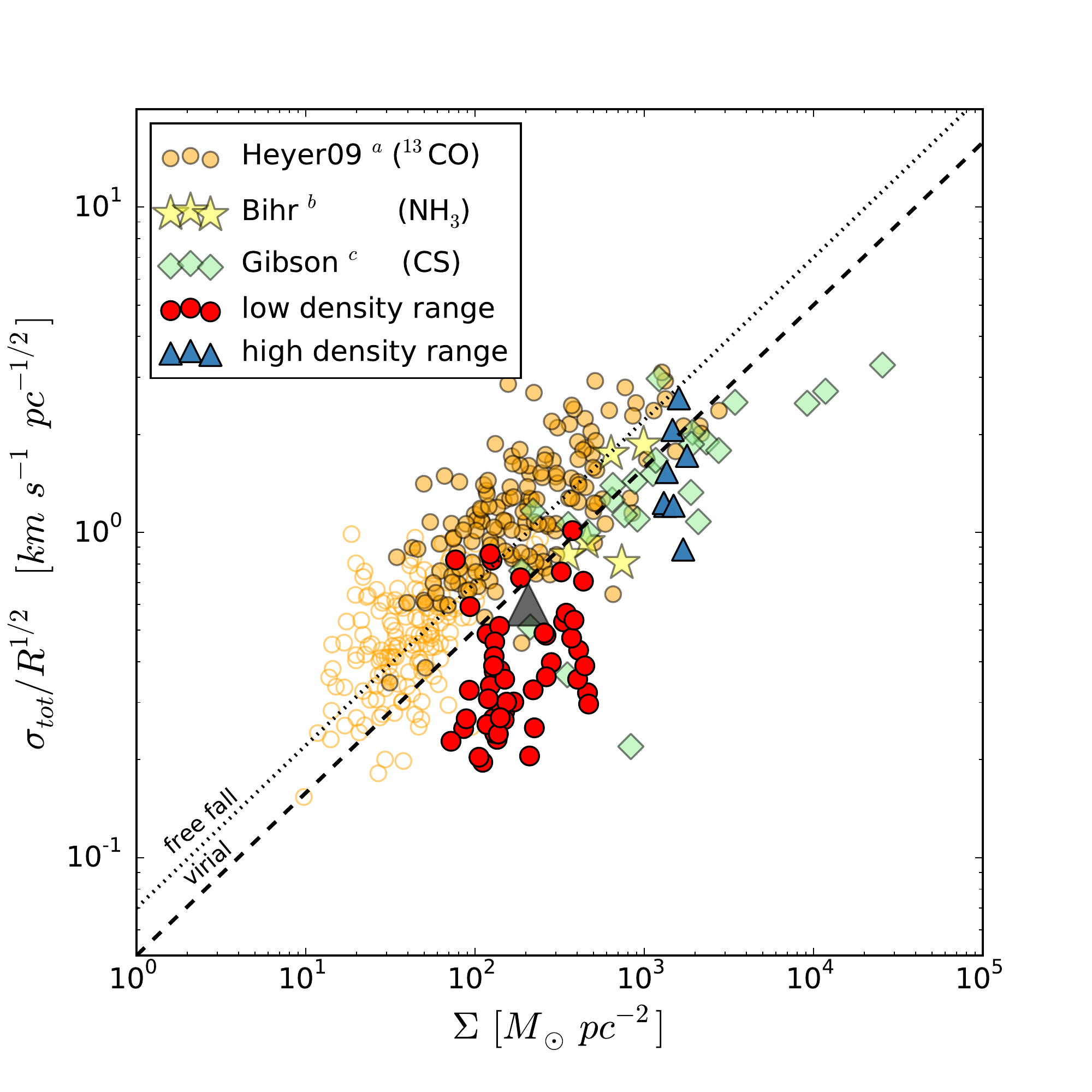}
\caption{Velocity dispersion-radius-surface density ($\sigma-R-\Sigma$) scaling relation for observations and simulations of MCs, clumps, and cores in the Galaxy.
Both plots show (orange open and filled circles) observations in $^{13}$CO reported by \citet{Heyer2009RE-EXAMININGCLOUDS};  (light green diamonds) Galactic infrared-dark clouds observed in CS \citep{Gibson2009MOLECULARCONDITIONS}; and (yellow stars) infrared dark clouds observed with NH$_{3}$ \citep{Bihr2015KinematicFormation}.
The black triangle shows the constant value of the column density reported by \citet{Solomon1987MassClouds} in their size-density relation. Both plots show the simulated objects captured by the low (red circles) and high (blue triangles) density ranges at (left) $t_{SG}=0$, and (right) after evolution.
Clumps are denoted with triangles because they are lower limits for the velocity dispersion, and upper limits for the cloud radius and surface density.
The dashed and dotted lines correspond to the relation $\sigma/R^{1/2} \propto \Sigma^{1/2}$, where the dashed line corresponds to the velocity dispersion for a uniform spherical cloud in virial equilibrium and the dotted line the apparent velocity dispersion for a cloud in free-fall collapse.}
\end{figure*} 

Since the early studies of the scaling relations in clouds using $^{12}$CO, new observations with a variety of tracer molecules sensitive to different density regimes have revealed that the scaling of the velocity dispersion not only depends on the radius of the cloud, $ \sigma \propto R^{1/2}$, but also varies systematically with the surface density of the cloud, $\sigma \propto R^{1/2} \Sigma^{1/2}$  \citep{Heyer2009RE-EXAMININGCLOUDS, Ballesteros-Paredes2011GravityRelation}.
L81 and \citet{Solomon1987MassClouds} were limited to observations of $^{12}$CO, exposing only low to intermediate density molecular gas, as discussed in Section \ref{sec:CloudCatalog}.
Because the column density traced by $^{12}$CO reaches the same maximum value in all but the smallest clouds, MCs traced by this species appear to all have almost the same column density, when in fact this is purely a radiative transfer effect \citep[e.g.][]{Ballesteros-Paredes2002PhysicalModels}.

Multitracer observations follow the dynamics of gas for a wide range of densities down to dense cores \citep{Caselli1995TheCores,Gibson2009MOLECULARCONDITIONS, Roman-Duval2011TurbulenceCalibration, Bihr2015KinematicFormation}.
\citet{Heyer2009RE-EXAMININGCLOUDS} re-examined the $\sigma-R$ relation for a subset of the $^{13}$CO GRS catalog, corresponding to the same clouds analyzed earlier in  $^{12}$CO by \citet{Solomon1987MassClouds}.
The new column densities and masses are directly calculated assuming local thermodynamical equilibrium (LTE), but without assumptions about the virial mass.
Given the higher resolution of the data and the usage of a more transparent tracer, $^{13}$CO, it is also possible to analyze denser cloud sub-structure.

Figure \ref{fig:Heyer-Plot} compares observations of velocity dispersion at widely varying cloud surface density and radius including GRS clouds \citep{Heyer2009RE-EXAMININGCLOUDS}, and infrared dark clouds \citep{Gibson2009MOLECULARCONDITIONS,Bihr2015KinematicFormation}, to our resolved cloud population at $t_{SG} = 0$ and to our evolved cloud population. 
As a reference, a black triangle indicates the median values of \citet{Solomon1987MassClouds} of $\sigma/R_{pc}^{1/2} = 0.72$ corresponding to a uniform surface density of $\Sigma = 206 \text{ M}_{\odot} \text{ pc}^{-2}$.

The simulated cloud population at t$_{SG} = 0$ exhibits low values of $\sigma/R^{1/2}$ and an anti-correlation with the cloud's mean surface density.
This happens because clouds formed during the non-self-gravitating evolution period of the simulations have very low velocity dispersions with respect to their masses and sizes.
The lack of correlation with $\sigma / R^{1/2}$ indicates that clouds formed in a non-self-gravitating, multi-phase, turbulent ISM have properties clearly inconsistent with observed MCs.  

On the other hand the evolved cloud and clump population is located near the expected region in the $\sigma-R-\Sigma$ parameter space, and has a slope consistent with the observed correlation, although the simulated clouds are systematically shifted to slightly lower values of $\sigma/R^{1/2}$ or higher values of $\Sigma$. 
The clumps show velocity dispersions, radii and surface densities similar to those predicted by \citet{Heyer2009RE-EXAMININGCLOUDS} relation. 
We caution that the properties of our evolved clump population do not fully resolve fragmentation, so they should be considered as upper limits on the cloud size and surface density.
Nevertheless, the clear correlation with the observations after self-gravity has been turned on, and not before, strongly suggests that it is gravitational contraction that dominates the observed velocity dispersions, rather than supernova driving alone.

\section{Discussion and Conclusions}
\label{sec:Discussion}

The balance between turbulent support and gravitational collapse has been argued to determine the formation and evolution of MCs \citep{MacLow2004ControlTurbulence}.
Simulations of isothermal turbulence continuously driven from large scales show that such turbulence can delay and inhibit star formation \citep{Klessen2000,Heitsch2001,Vazquez-semadeni2005,Federrath2013}.
The observed velocity dispersion-size relation has been suggested to originate from the inertial turbulent cascade with no dependence on the gas self-gravity \citep{Kritsuk2013ALaws, Padoan2015OriginTurbulence}. 
Observations show that the observed turbulent motions are dominated by the largest-scale modes \citep{MacLow2000CharacterizingTurbulence, Brunt2003LargeScaleClouds, Brunt2009TurbulentClouds}. 
However, no mechanism has yet been positively identified to continuously drive such large-scale turbulence in MCs.

The most viable candidate for maintaining diffuse ISM turbulence appears to be a combination of field SN explosions and superbubbles \citep{MacLow2004ControlTurbulence,Tamburro2009, Padoan2015OriginTurbulence} and accretion onto the galactic disk \citep{Klessen2010Accretion-drivenDisks, Klessen2014PhysicalMedium}.
However the results we have presented in section \ref{sec:sigma-R} show that SN explosions seem unable to drive turbulence within dense clouds and thus appear unlikely to be responsible for the observed velocity dispersion-size relation in MCs.

In our simulations prior to the onset of self-gravity, clouds form at the stagnation points of convergent flows driven by SN remnant and superbubble expansion.
Because of radiative cooling during their formation, this leaves them at lower temperatures and higher densities than their surroundings.
During this non-self-gravitating evolution the simulated clouds live very long lives, lasting tens to hundreds of megayears, enough time for the internal turbulence to decay, $t_{decay} \sim R / \sigma$  \citep{MacLow1998KineticClouds, Stone1998DissipationTurbulence, MacLow1999TheClouds}.
Although the clouds are constantly being deformed by SN explosions, these do not drive substantial internal turbulence.
While SN continue to explode in the diffuse ISM, the clouds maintain low velocity dispersions, which appears to be all that can be driven by the external turbulence.

A possible explanation for this behavior is that turbulence in the diffuse ISM has to climb up a gradient of several orders of magnitude in density to drive turbulent motions in the MC.
Momentum is conserved, though the energy drops due to radiative cooling.
As a result of momentum conservation, the velocity drops as the density increases, so that the turbulent motions in the resolved dense interior of the MCs remain well below a kilometer per second, more than an order of magnitude below the tens of kilometer per second driving flows.   
During the self-gravitating evolution of the simulation, clouds also seem to accrete material from their environment. 
We do not distinguish here between accretion driven turbulence and contraction.

Results presented here are in direct contradiction with the argument of \citet[hereafter P15]{Padoan2015OriginTurbulence} who suggest that SN explosions alone are responsible for the fast turbulent motions inside dense clouds, and also with the argument of \citet[hereafter K13]{Kritsuk2013ALaws}, who suggest that internal turbulent motions in molecular clouds originate by a supersonic turbulent cascade whether or not self-gravity is included.

It is difficult to directly compare the simulations presented in this work and those analyzed by both P15 and K13, as they differ in several critical characteristics.
P15 simulate a 250~pc$^{3}$ cubic, periodic, unstratified box with a minimum resolution $\Delta x = 0.24$~pc, as opposed to our stratified box, with a minimum resolution of $\Delta x = 0.95$~pc.  They show neither a $\sigma-R$ relation plot before self-gravity is turned on, nor the structure function of the same cloud before and after self-gravity.  Thus it remains unclear whether their suggestion that supernova driving dominates over self-gravity is well supported. 

It is possible that numerical dissipation in our lower resolution models suppresses velocity dispersion at small scales. However, we demonstrated in Section \ref{sec:Resolution Study} that the velocity dispersions for clouds larger than 4.8~pc do not change for resolution down to $\Delta x = 0.47$~pc. 

In addition, close inspection of Figure 3 of P15 supports our interpretation of gravitational contraction driving turbulent motions in dense gas with $n>100$~cm$^{-3}$. Before the onset of self-gravity at 45 Myr, the mean kinetic energy is roughly constant, with irregular peaks probably corresponding to the formation of dense structures in convergent flows, that quickly decay after $\sim$1~Myr. After the onset of self-gravity, it appears that the mean kinetic energy increases as a function of time.  This behaviour can also be observed in their Figure 4, where before self-gravity is active, the mean velocity dispersion is on average 5--6 km~s$^{-1}$. However after the onset of self-gravity, the mean velocity dispersion clearly increases with time up to an average value of 15--16 km~s$^{-1}$. We believe this is at least in part due to gravitational collapse and not solely to SN explosions.

K13 use a suite of periodic-box simulations to argue for a supersonic-turbulence origin of Larson's laws. 
In all of the K13 simulations, turbulence is driven by large scale forcing and the analysis is performed once a steady state is reached.
This is a crucial difference between our setup and that of K13,  as large scale forcing acts as a volume force term instead of the surface force that one expects from SN-driven or accretion-driven turbulence. 
Only one of K13's simulations allows for turbulence to decay and includes gas self gravity (HD3, \citet{Kritsuk2011OnClouds}).
This simulation is evolved only for a fraction of a free fall time in the presence of self-gravity, $0.43 t_{{\rm ff}}$, which also corresponds to a small fraction of the dynamical crossing time, $0.23 t_{{\rm dyn}}$.
Thus, too little time has elapsed for the kinetic energy of the steady state to decay and for self-gravity to affect the velocity structure of the clouds.

For the simulations presented in this work, when self-gravity is turned on, all the clouds begin to collapse simultaneously.
This scenario is, of course, only a crude approximation for the evolution of MCs in the Galaxy, as gas self-gravity is always present during the formation and evolution of the clouds, while stellar feedback quickly sets in, preventing long-lived quiescent clouds from occurring.
Thus, we do not actually expect that clouds go through a phase of low velocity dispersion, as observed in our clouds at t$_{SG} = 0$, but rather expect the ensemble of observable clouds to always have velocity dispersions consistent with Larson's relation.
It is also important to take into account the time it takes to build a sufficient amount of CO to be detectable in a cloud. 
Colliding flow simulations including non-equilibrium chemistry show that there is a long (up to 10 Myr) phase during which the cloud is held together by ram pressure. During this phase the cloud has enough density to form H$_{2}$, but not enough dust extinction to form CO.
Once the cloud becomes Jeans unstable and begins contracting due to its own self-gravity, it reaches column densities sufficient for CO to be shielded and abundant enough to be observed \citep{Clark2012HowCloud, Clark2014ColumnDensSF}.
This suggests that clouds observed with CO emission are always in a state of gravitational contraction, giving rise to Larson's relations.

As clouds form and turbulent velocities decay, the clouds become more and more self-gravitating. Localized centers of gravitational collapse accelerate the gas, producing a chaotic set of supersonic motions easily interpreted as being due to supersonic turbulence.  Given the high Reynolds numbers prevalent in this system, the motions likely are indeed turbulent, but driven primarily by hierarchical gravitational collapse.

The results presented here strongly contradict the hypothesis that SN explosions alone can drive turbulence in MCs that reproduces the velocity dispersion-radius relation or its surface density dependent corollary. 
Only when self-gravity is included do the velocity dispersions in the simulated clouds increase to values in agreement with observations, as proposed by \citet{Ballesteros-Paredes2011GravityRelation}, and agreeing with the more general proposal by \citet{Klessen2010Accretion-drivenDisks}.

Supernova-driven turbulence remains essential in driving the non-linear density fluctuations that provide the seeds for hierarchical collapse to proceed.  
This is seen in Figure \ref{fig:StratBox_composite} where the close-up image clearly shows that MCs are far from uniform spheres, but rather have complex, filamentary shapes and density distributions.
In this cloud, gravitational collapse does not proceed uniformly but rather hierarchically, depending on the local density distribution. Our results thus support the hypothesis that global collapse of hierarchically structured clouds drives the non-thermal motions observed inside MCs.

Our simulations neglect any explicit correlation between the location of SN explosions and the position of the parent clouds of clusters.  Simulations of SN feedback in periodic boxes, have shown that the ISM structure is strongly dependent on the location of the SN explosions, whether explosions are correlated with density peaks, randomly distributed, or something in between \citep{Gatto2015,Li2015SupernovaMedium}.
However observations demonstrate that only $25\%$ of identified SN remnants are superposed on detectable molecular hydrogen emission \citep{Froebrich15}, while only 15\% show direct maser evidence of interaction with molecular gas \citep{Hewitt2009}. 
Furthermore, studies of molecular cloud disruption suggest that ionizing radiation has substantially greater effect than winds or SNe \citep{Rogers2013,Dale2014,Walch2015}.
For a more realistic study of the correlation of the SN explosions with respect to the parent cloud, though, models of self-consistent star formation and feedback from massive stars will be required, which we are currently pursuing.

When self-gravity is activated in our simulations, clouds quickly begin to collapse. 
This means that the clouds in our simulations are not supported by magnetic, thermal, or turbulent pressure.
Collapsing clouds increase their internal velocity dispersion as gravitational potential energy is converted to kinetic energy.
After a free fall time, clouds approximate equipartition, $|E_{g}| \sim E_{k}$ and evolve in that state from there on. 
However, it needs to be emphasized that equipartition does not imply virial equilibrium, but instead just means that the cloud is converting potential into kinetic energy as it collapses, so that both should be comparable \citep{Ballesteros-Paredes2006}.
Equipartition velocity dispersions are similar to those predicted for clouds in equilibrium, as clouds in equilibrium should also have kinetic energies comparable to the cloud's gravitational potential energy.
It is for this reason that it is so difficult to differentiate between collapsing clouds and clouds in equilibrium.

The collapse of a hierarchically structured cloud will proceed at different speeds in different parts of the cloud, since higher density cores have shorter free fall times than their envelopes \citep{Elmegreen2007}.
This idea corresponds to the scenario outlined by \citet{Heyer2009RE-EXAMININGCLOUDS}, and \citet{Ballesteros-Paredes2011GravityRelation} where they discuss the dependence of the velocity dispersion not only on the cloud's size, but on the surface density as well.
This scenario agrees with models presented by \citet{Elmegreen1993}, \citet{BallesterosParedes1999a,  BallesterosParedes1999b}, \citet{Hartmann2001}, \citet{Vazquez-Semadeni2003, VazquezSemadeni2006MolecularFormation}, and \citet{Heitsch2005, Heitsch2006}, where clouds never reach a state of virial equilibrium, but instead are in a constant state of evolution and collapse.

We speculate that our results support the original hypothesis that MCs are generally collapsing suggested by \citet{Goldreich1974MolecularClouds}, but with a twist to the objection by \citet{Zuckerman1974RadioMolecules} that the free fall collapse of all the molecular gas in the Galaxy would result in far too high a star formation rate.
While the clouds are in a state of collapse, they do not collapse globally but in a hierarchical fashion. 
Before clouds can collapse as a whole and transform most of their mass into stars, dense regions collapse first, forming stars early in the cloud's life \citep{VazquezSemadeni2006MolecularFormation,VazquezSemadeni2007MolecularConditions,Elmegreen2007}.
Once star formation in the cloud begins, stellar feedback can disrupt the cloud, maintaining a low star formation efficiency for the MC as a whole.

\section{Summary}
\label{sec:Conclusions}

We present numerical simulations of a stratified, multiphase, magnetized, SN-driven, turbulent ISM.
We measure the properties of the cloud population that form in this turbulent ISM at an arbitrary time in the simulation prior to including self gravity. We then include gas self-gravity, measure the properties of the cloud population at different evolutionary stages and compare them with observations, focusing in particular on the relations between velocity dispersion, radius, and column density. We find: 

\begin{itemize}

\item SN feedback in the diffuse ISM only appears able to drive turbulent motions in dense MCs under a kilometer per second, inconsistent with observations. This is most likely because momentum conservation allows only the fast flows in the diffuse medium to drive turbulent velocities in the dense MCs slower by a factor of the density contrast \citep{Klessen2010Accretion-drivenDisks}.
\item MCs and their major internal substructures continuously contract gravitationally. We find no evidence for static clouds or clumps in equilibrium. Our simulations include magnetic fields, but these also cannot prevent contraction.       
 \item Gravitational contraction thus appears most likely to be the origin of the velocity dispersion-size relation, driving non-thermal motions \citep{Traficante2015} correlated with the cloud size as observed $\sigma_{tot} \propto R^{1/2}$ \citep{Larson1981,Solomon1987MassClouds,Falgarone2009IntermittencyScale}.  
\item Clouds are in a state of hierarchical contraction, where the velocity dispersion of a cloud or a clump depends not only on the size, but also on the column density, $\sigma^2 \propto R \Sigma$ \citep{Heyer2009RE-EXAMININGCLOUDS,Ballesteros-Paredes2011GravityRelation}.
\end{itemize}

\acknowledgments{
  Thanks to Cara Battersby, Javier Ballesteros-Paredes, L\'aszl\'o Sz\H{u}cs, Jens Kauffmann, Rahul Shetty, Andrea Gatto, Alex Hill, Simon Glover, Paul Clark, Rowan Smith and Enrique V\'azquez-Semadeni for useful discussions.
  This work was supported by NSF grant AST11-09395.  M-MML was additionally supported by the Alexander-von-Humboldt Stiftung. Computations were performed on TACC Stampede under grant TG-MCA99S024 from the Extreme Science and Engineering Discovery Environment (XSEDE), which is supported by NSF grant OCI-1053575. 
RSK acknowledges support from the Deutsche Forschungsgemeinschaft (DFG) via SFB 881 "The Milky Way System" (sub-projects B1, B2 and B8), and SPP 1573 "Physics of the ISM". Furthermore RSK thanks the European Research Council for funding under the European Community’s Seventh Framework Programme via the ERC Advanced Grant "STARLIGHT" (project number 339177).
Visualizations were performed with {\tt yt} \citep{Turk2011AData}. }

\bibliographystyle{apj}
\bibliography{Mendeley}

\end{document}